\documentclass{ws-procs9x6}            

\def\be{\begin{equation}}
\def\ee{\end{equation}}
\def\ber{\begin{eqnarray}}
\def\eer{\end{eqnarray}}

\def\rv{{\bf r}}

\def\pv{{\bf p}}

\def\qv{{\bf q}}

\def\ep{{\epsilon}}
\def\br{{\bf r}}

\def\bp{{\bf p}}

\def\bq{{\bf q}}

\def\nn{\nonumber}
\def\nn{\nonumber}

\def\TG{\tilde{G}}

\def\bsigma{\boldsymbol{\sigma}}

\def\be{\begin{equation}}
\def\ee{\end{equation}}
\def\ber{\begin{eqnarray}}
\def\eer{\end{eqnarray}}
\def\nn{\nonumber}
\newcommand{\agrave}{\`a}

\begin{document}
\title{ Spin-charge coupling effects in a two-dimensional electron gas}

\author{Roberto Raimondi}

\address{Dipartimento di Matematica e Fisica, Universit{\agrave} Roma Tre,\\
Via della Vasca Navale 84, Rome, 00146, Italy\\
E-mail: roberto.raimondi@uniroma3.it\\
www.uniroma3.it}

\author{Cosimo Gorini}

\address{Institut f\"ur Theoretische Physik, Universit\"at Regensburg,\\
93040 Regensburg, Germany\\
E-mail: cosimo.gorini@physik.uni-regensburg.de}

\author{Sebastian T\"olle}

\address{Institut f\"ur Physik, Universit\"at Augsburg\\
Universit\"atsstr. 1, 86135 Augsburg, Germany\\
E-mail: sebastian.toelle@physik.uni-augsburg.de}

\begin{abstract}
In these lecture notes we study the disordered two-dimensional electron gas in the presence of Rashba spin-orbit coupling, 
by using the Keldysh non-equilibrium Green function technique.  We  describe
the effects of the spin-orbit coupling in terms of a SU(2) gauge field and derive
a generalized Boltzmann equation for the charge and spin distribution functions.
We then apply the formalism to discuss the spin Hall and the inverse spin galvanic (Edelstein) effects.  
Successively we show how to include, within the generalized Boltzmann equation, the side jump, the skew scattering and the spin current swapping processes originating from the extrinsic spin-orbit coupling due to impurity scattering.
\end{abstract}

\keywords{Spin-orbit coupling; Electronic transport; Many-body Green function; Disordered systems.}

\bodymatter

\section{Introduction}\label{sec1}

These lecture notes are based mainly on the work by Gorini et al.  of Ref.\citenum{GoriniPRB10}, where by means of a gradient expansion a generalized Boltzmann equation
with SU(2) gauge fields was obtained for the disordered Rashba model.  The inclusion of the extrinsic spin-orbit coupling (SOC) from impurities 
in the SU(2) formalism was later considered in the work by Raimondi et al. in Ref.\citenum{Raimondi_AnnPhys12}.    Hence, the aim of these lecture notes is to provide a self-contained 
and pedagogical introduction to the disordered two-dimensional electron gas (2DEG) with both  intrinsic (Rashba) and extrinsic SOC within the SU(2) gauge-field approach. 
The lecture notes by Tatara in this series are a good complementary reading dealing with electron transport in ferromagnetic metals\cite{Tatara2016}.

The layout of these lecture notes is the following. In Section 2 we write down the quantum kinetic equation for 
the fermion Green function in the presence of U(1), associated with the electromagnetic field, and SU(2) gauge fields.  
The standard model of disorder is introduced in Section 3.  Whereas in Section 2 we derive the hydrodynamic SU(2) derivative 
of the Boltzmann equation, in Section 3 we obtain an expression for the collision integral describing the scattering from impurities. 
In Section 4 we apply the formalism to the disordered Rashba model and derive the Bloch equation for the spin density, describe the Dyakonov-Perel spin relaxation, and discuss a thermally induced spin polarization. In Section 5 we introduce SOC from impurity scattering and analyze the so-called side jump mechanism, which manifests as a correction to the velocity operator. In Section 6 we discuss the skew scattering mechanism. Both intrinsic and extrinsic SOCs contribute to the spin relaxation.  
Extrinsic SOC gives rise to Elliott-Yafet spin relaxation, which is covered in Section 7, together with the complete form of the Boltzmann equation. Section 8 states our conclusions.  Throughout we use units such that $\hbar=c=1$.

\section{The kinetic equation and  the SU(2) covariant Green function}\label{sec2}

We begin by defining the Keldysh Green function (for an introduction see, e.g., the book by Rammer \cite{Rammer_qkt})
\be
\check G =\begin{pmatrix}
  G^R    &  G^K  \\
   0   &  G^A
\end{pmatrix},
\label{eq_1}
\ee
where the retarded $G^R$, Keldysh $G^K$ and advanced $G^A$ components are given by
\ber
G^R(\rv_1,t_1;\rv_2,t_2)&=&-i\Theta(t_1-t_2)\langle \psi(\rv_1,t_1)\psi^{\dagger}(\rv_2,t_2)+\psi^{\dagger}(\rv_2,t_2)\psi(\rv_1,t_1)\rangle\nn\\
G^K(\rv_1,t_1;\rv_2,t_2)&=&-i\langle \psi(\rv_1,t_1)\psi^{\dagger}(\rv_2,t_2)-\psi^{\dagger}(\rv_2,t_2)\psi(\rv_1,t_1)\rangle\nn\\
G^A(\rv_1,t_1;\rv_2,t_2)&=&i\Theta(t_2-t_1)\langle \psi(\rv_1,t_1)\psi^{\dagger}(\rv_2,t_2)+\psi^{\dagger}(\rv_2,t_2)\psi(\rv_1,t_1)\rangle.\nn
\eer
In the above definitions $\psi(\rv_1,t_1)$ and $\psi^{\dagger}(\rv_2,t_2)$ are Heisenberg field operators for fermions and $\Theta (t)$ the Heaviside step function. In the following we will be  concerned with spin one-half  fermions. As a consequence, all entries of $\check G$ become two by two matrices.

To derive a kinetic equation, it is useful to introduce Wigner mixed  coordinates. To this end we perform a Fourier transform with respect to both space ($\rv_1-\rv_2$) and time ($t_1-t_2$) relative coordinates
\be
\check G (\pv,\ep; \rv,t)=\int {\rm d} (t_1-t_2) \int {\rm d}(\rv_1 -\rv_2) \check G(\rv_1,t_1;\rv_2,t_2) e^{i\left[\ep (t_1-t_2)-\pv \cdot (\rv_1-\rv_2)\right]}.
\label{eq_2}
\ee

The first step in the standard derivation of the kinetic equation is the left-right subtracted Dyson equation
\be
\left[  \check G_0^{-1}(x_1,x_3)\overset{\otimes}{,} \check G(x_3,x_2) \right]=0,
\label{eq_3}
\ee
where we have used space-time coordinates $x_1\equiv (\rv_1,t_1)$ {\it etc}. In \eref{eq_3},
the symbol $\otimes$ implies integration over $x_3$ and matrix multiplication both in Keldysh and spin (if any) spaces. Furthermore
\be
 \check G_0^{-1}(x_1,x_3)
=\left( i\partial_{t_1}-H \right)\delta (x_1-x_3),
\label{eq_4}
\ee
where $H$ is the Hamiltonian operator.  In these lecture notes we do not consider electron-electron interaction. Quite generally the Hamiltonian operator takes the form
\be
H=\frac{(-i\nabla_{\rv}  +e{\bf A}(\rv,t))^2}{2m}-e\Phi (\rv,t)+V(\rv).
\label{eq_5}
\ee
Here $e=|e|$ and we have assumed negatively charged particles.
In \eref{eq_5} the scalar and vector potential have a two by two matrix structure, which can be shown by expanding them in the basis of the Pauli matrices
\be
\Phi =\Phi^0 \sigma^0 +\Phi^a \frac{\sigma^a}{2}, \
{\bf A} ={\bf A}^0 \sigma^0 +{\bf A}^a \frac{\sigma^a}{2}, \ \ a=x,y,z,
\label{eq_6}
\ee
and summation over the repeated indices is understood.
The $\sigma^0$-components are the electromagnetic scalar and vector potentials associated with the U(1) gauge invariance.
$V(\rv)$ describes the disorder potential due to impurities and defects. In this section we set $V(\rv)=0$ and postpone its discussion to the following section. 
The $\sigma^a$-components are an SU(2) gauge field, whose scalar and vector components can be used to respectively
describe a Zeeman/exchange term and SOC -- in our case Rashba SOC, as will be shown in  \sref{sec3a}. 
For the time being, we do not consider a specific form of the SU(2) gauge field $(\Phi^a, {\bf A}^a)$.

The goal of a kinetic equation is to describe non-equilibrium phenomena. In general this is a formidable task. 
However, for close-to-equilibrium phenomena or for non-equilibrium ones occurring on scales large compared to microscopic ones, 
it is possible to derive an effective kinetic equation by means of the so-called gradient expansion.  
The idea is based on the observation that under specific circumstances the Green function varies fast with respect to the relative coordinate $x_1-x_2$ and much more slowly with respect to the center-of-mass one $(x_1+x_2)/2$.  In equilibrium, for a translationally invariant system, the Green function does not depend on the center-of-mass coordinate at all.

To understand how the gradient expansion works, consider the convolution of two quantities
\be
(A\otimes B)(x_1,x_2) = \int {\rm d} x_3 A(x_1,x_3)B(x_3,x_2),
\nn
\ee
which can be equivalently expressed 
as a function of center-of-mass and relative coordinates
\be
\int {\rm d} x_3 A\left( \frac{x_1+x_3}{2},x_1-x_3\right)
B\left( \frac{x_3+x_2}{2},x_3-x_2\right).
\nn
\ee
 Next, replace $x_1+x_3=x_1+x_2+x_3-x_2$ and $x_3+x_2=x_1+x_2-(x_1-x_3)$ in the first argument of $A$ and $B$, respectively. By Taylor expanding $A$ with respect to $x_3-x_2$
 in its first argument and $B$ with respect to $x_1-x_3$ in its first argument, after Fourier transforming according to \eref{eq_2},
one gets
 \be
 A(x,p)B(x,p)+
 \frac{i}{2}\big(\partial_{\mu} A(x,p)\big) \big(\partial_p^{\mu} B(x,p) \big)-
 \frac{i}{2}\big(\partial_p^{\mu} A(x,p)\big) \big(\partial_{\mu} B(x,p)\big),
 \label{eq_8}
 \ee
where we have introduced the compact (relativistic)  space-time notations
\be
x^{\mu}=(t,\rv), \ x_{\mu}=(-t,\rv), \ p^{\mu}=(\ep,\pv), \ p_{\mu}=(-\ep,\pv)
\label{eq_9}
\ee
and
\be
\partial^{\mu}\equiv \frac{\partial}{\partial x_{\mu}}, \
\partial_{\mu}\equiv \frac{\partial}{\partial x^{\mu}}, \
\partial_p^{\mu}\equiv \frac{\partial}{\partial p_{\mu}}, \
\partial_{p,\mu}\equiv \frac{\partial}{\partial p^{\mu}}
\label{eq_10}
\ee
in a such a way that the product $p^{\mu}x_{\mu}=-\ep t +  \pv\cdot \rv$ has the correct Lorentz metrics. \Eref{eq_3} acquires then the form
\be
-i \left[ \check G_0^{-1}, \check G\right]
+\frac{1}{2}
\left\{\big( \partial^{\mu} \check G_0^{-1} \big) , \big(\partial_{p,\mu}\check G\big)\right\}
-\frac{1}{2}
\left\{ \big(\partial_p^{\mu} \check G_0^{-1}\big) , \big(\partial_{\mu}\check G\big)\right\}
=0.
\label{eq_11}
\ee

The Hamiltonian (\ref{eq_5}) is invariant under a gauge transformation $O(x)$, which locally rotates the spinor field
\be
\psi' (x)=O(x)\psi (x), \ \psi'^{\dagger}(x)=\psi^{\dagger}(x)O^{\dagger}(x), \  O(x) O^{\dagger}(x)=1.
\label{eq_12}
\ee
The Green function, however, is not locally covariant, i.e. its transformation depends on two distinct space-time points
\be
\check G (x_1,x_2)\rightarrow O(x_1) \check G(x_1,x_2) O^{\dagger}(x_2).
\label{eq_13}
\ee
Physical observables, which are locally covariant, are obtained by considering the Green function in the limit of coinciding space-time points. It is then useful to introduce a {\it locally} covariant Green function
\be
 \check {\tilde G}(x_1,x_2)=U_{\Gamma}(x,x_1)\check G(x_1,x_2)U_{\Gamma}(x_2,x)
\label{eq_14}
\ee
where 
\be
U_{\Gamma}(x,x_1)={\cal P}\exp \left(- i \int_{x_1}^x eA^{\mu}(y){\rm d}y_{\mu}\right).
\label{eq_15}
\ee
The line integral of the gauge field is referred to as the Wilson line. In \eref{eq_15} ${\cal P}$ is a path-ordering operator and $A^{\mu}=(\Phi, {\bf A})$, $A_{\mu}=(-\Phi, {\bf A})$. Since the Wilson line transforms covariantly
 \be
 U_{\Gamma}(x,x_1)\rightarrow O(x)U_{\Gamma}(x,x_1) O^{\dagger}(x_1),
 \label{eq_16}
 \ee
one easily sees that the covariant Green function $\check{\tilde G}$ transforms in a locally covariant way
\be
\check{\tilde G}(x_1,x_2)\rightarrow O(x) \check{\tilde G}(x_1,x_2) O^{\dagger}(x).
\label{eq_17}
\ee
When $x_1=x_2=x$, the locally covariant Green function coincides with the original Green function. By inverting \eref{eq_14}, one can,  via \eref{eq_11}, obtain an equation for the locally covariant Green function. Due to the non-Abelian character of the gauge field, \eref{eq_14} is not easy to handle. In the spirit of the gradient approximation, we assume that 
$\partial^\mu \partial_{\mu,p} \ll 1$. In addition we also assume that $eA^\mu\partial_{\mu,p}\ll 1$. This assumption can be justified on physical grounds once an explicit form is assigned to $A$.
Under these assumptions, \eref{eq_14} becomes
\be
\check{\tilde G}=\check G-\frac{1}{2} \{ eA^{\mu}\partial_{p,\mu},\check G\}
\label{eq_18}
\ee
and its inverse
\be
\check G=\check{\tilde G}+\frac{1}{2} \{e A^{\mu}\partial_{p,\mu},\check{\tilde G}\}.
\label{eq_19}
\ee
By using the decomposition 
\be
\delta (x_1-x_2)=\int \frac{{\rm d}^{d+1}p}{(2\pi)^{d+1}}e^{ip^{\mu}(x_{1,\mu}-x_{2,\mu})},
\label{eq_fourier}
\ee
one obtains
\be
\check G_0^{-1}(x,p)=\ep -\frac{(\pv+e{\bf A}(x))^2}{2m}+e\Phi (x)=\ep-\frac{p^2}{2m} - V^{\mu}e A_{\mu} - \frac{e^2{\bf A}^2}{2m},
\label{eq_20}
\ee
from which
\ber
\partial_p^{\mu}\check G_0^{-1}(x,p) &=& -V^{\mu} - \left( \partial_p^\mu V^\nu  \right) e A_\nu, \nn \\ 
\partial^{\mu}\check G_0^{-1}(x,p) &=& -V^{\nu}\partial^{\mu}e A_{\nu}.
\label{eq_21}
\eer 
In the above  $V^{\mu}=(1, \pv /m)$ is  the $d$-current operator, $d$ being the space dimensionality. In the second equation of (\ref{eq_21}) we neglected  the term $\partial^\mu {e^2\bf A}^2 /{2m}=e{\bf A}\cdot \partial^{\mu} e{\bf A}$  because it gives a small correction to $\pv\cdot \partial^{\mu} e{\bf A}$ when $\pv \sim \pv_F$.
Performing the {\it shift} transformation of \eref{eq_18} in \eref{eq_20} gives
\be
\check{\tilde G}_0^{-1}=\ep -\frac{p^2}{2m}.
\label{eq_22}
\ee

We have now all the necessary  ingredients to obtain the equation for $\check{\tilde G}$. We begin by considering the first term of \eref{eq_11}. By applying the {\it shift} transformation of \eref{eq_18} and expressing $\check G$ in terms of $\check{\tilde G}$ via \eref{eq_19}, we obtain
\be
-i\left( \left[\check G_0^{-1}, \check{\tilde G} \right]-\frac{1}{2}\Big\{eA_{\mu},\partial_p^\mu \left[\check G_0^{-1}, \check{\tilde G} \right] \Big\}+\frac{1}{2}\left[ \check G_0^{-1}, \Big\{eA_{\mu},\partial_{p}^\mu \check{\tilde G} \Big\}\right]\right).
\label{eq_23}
\ee
By means of the identity $\{A,\left[B,C\right]\}-\left[B,\{A,C\}\right]=\{\left[A,B\right],C\}$
we get
\be
ieV^{\mu} \left( \left[A_{\mu}, \check{\tilde G} \right]+\frac{1}{2}\Big\{e \left[A_{\mu}, A_{\nu}\right],\partial_{p}^\nu \check{\tilde G}\Big\}\right).
\label{eq_24}
\ee
As for the second term of \eref{eq_11}
\be 
\frac{1}{2}
\left\{\big( \partial^{\mu} \check G_0^{-1} \big) , \big(\partial_{p,\mu}\check G\big)\right\} \rightarrow
- \frac{e}{2}V^{\mu}\Big\{ \left(\partial^{\nu}A_{\mu}\right),\partial_{p,\nu}\check{\tilde G}\Big\},
\label{eq_26}
\ee
where the last step  follows by considering the first order of the gradient expansion.
Finally, for the last term of \eref{eq_11}
\be
-\frac{1}{2}
\left\{ \big(\partial_p^{\mu} \check G_0^{-1}\big) , \big(\partial_{\mu}\check G\big)\right\} \rightarrow V^{\mu} \left[\partial_{\mu}\check{\tilde G}+\frac{1}{2}\Big\{\left(e\partial_{\mu}A^{\nu} \right),\partial_{p,\nu}  \check{\tilde G}\Big\} \right],
\label{eq_25}
\ee
where we omitted terms $\sim \partial_{\mu}\partial_p^\nu\check{\tilde G}$ within the first order accuracy  of the gradient expansion.
By collecting the results of Eqs.(\ref{eq_24}-\ref{eq_25}), the equation for  $\check{\tilde G}$ reads
\be
V^{\mu} \left[\tilde\partial_{\mu}\check{\tilde G}+\frac{1}{2}\Big\{eF_{\mu\nu},\partial_{p}^\nu\check{\tilde G}\Big\} \right]=0, \label{eq_27}
\ee
where we have introduced the covariant derivative
\be
\tilde\partial_{\mu}\check{\tilde G}=\partial_{\mu}\check{\tilde G} + i\left[e A_{\mu}, \check{\tilde G}\right]
\label{eq_28}
\ee
and the field strength
\be
F_{\mu\nu}=\partial_{\mu}A_{\nu}-\partial_{\nu}A_{\mu} +
i e \left[ A_{\mu},A_{\nu}\right].
\label{eq_29}
\ee
It is useful at this stage to separate the space and time parts and rewrite \eref{eq_27} as
\be
\left(\tilde\partial_t +\frac{\pv}{m}\cdot \tilde \nabla_{\rv}\right)\check{\tilde G}-
\frac{e}{2}\Big\{ \frac{\pv}{m}\cdot{\bf E},\partial_{\ep} \check{\tilde G}\Big\}+
\frac{1}{2}\Big\{ {\bf F},\nabla_{\pv} \check{\tilde G}\Big\}=0,
\label{eq_30}
\ee
where the generalized Lorentz force reads
\be
{\bf F} = - e\left( {\bf E} + \frac{\bp}{m} \times {\bf B} \right)
\label{eq_31}
\ee
with the U(1)$\times$SU(2) fields given by
\ber
{\bf E} &=& -\partial_t {\bf A}-\nabla_{\rv} \Phi +i e\left[ \Phi, {\bf A}\right] , \nn \\
B_i &=& \frac{1}{2} \varepsilon_{ijk} F^{jk} .
\label{eq_32}
\eer
\Eref{eq_30} is the quantum kinetic equation.  One can integrate over the energy $\ep$,
corresponding to the equal-time limit, in order to obtain a semiclassical kinetic equation.
We define the distribution function as
\be
f(\pv, \rv, t)\equiv \frac{1}{2}\left[1+\int \frac{{\rm d}\ep}{2\pi i} \tilde G^K (\pv, \ep;\rv, t) \right],
\label{eq_33}
\ee
which is a matrix in spin space, $f = f^0 \sigma^0 + f^a \sigma^a,\,a=x,y,z$.
By taking the Keldysh component of \eref{eq_30} we get
\be
\left(\tilde\partial_t +\frac{\pv}{m}\cdot \tilde \nabla_{\rv}\right)f(\pv, \rv, t)
+
\frac{1}{2}\Big\{ {\bf F}\cdot \nabla_{\pv}, f(\pv, \rv, t)\Big\}=0.
\label{eq_34}
\ee
We have then obtained a  generalization of the Boltzmann equation, where space and time derivatives are replaced by  the covariant ones and the  generalized Lorentz force appears. We may then introduce the  density and current by integrating over the momentum
\be
\rho (\rv, t)=\sum_{\pv} f(\pv, \rv, t), \ 
{\bf J}(\rv,t)=\sum_{\pv}\frac{\pv }{m} f(\pv, \rv, t).
\label{eq_35}
\ee
Hence the integration over the momentum  of \eref{eq_34} leads to a continuity-like equation
\be
\tilde\partial_t \rho (\rv,t)+\tilde\nabla_{\rv}\cdot {\bf J}(\rv,t)=0.
\label{eq_36}
\ee
We will use the above equation in Section \ref{sec3a}, when discussing the spin Hall and inverse spin galvanic/Edelstein
 effects in the Rashba model.

\section{The standard model of disorder and the diffusive approximation}
\label{sec3}

\begin{figure}
\includegraphics[width=3.in]{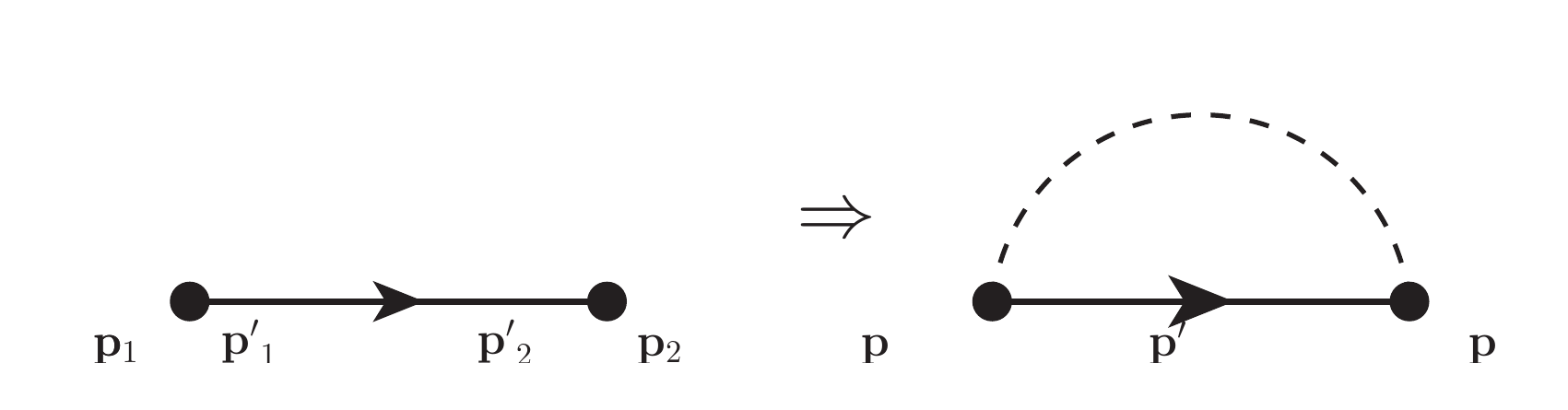}
\caption{Self-energy diagram to second order in the impurity potential (black dot vertex). The diagram on the left is before the impurity average, which is carried in the diagram on the  right  as a dashed line connecting the two impurity insertions. Notice that the impurity average in momentum space yields $\langle V(\pv_1 -\pv'_1)V(\pv'_2-\pv_2)\rangle =n_iv_0^2\delta (\pv_1 -\pv'_1+\pv'_2-\pv_2)$.}
\label{fig1}
\end{figure}

In this section we consider the effect of disorder  due to impurity scattering. According to the standard model of disorder\cite{AGD1975}
the potential $V(\rv)$ is assumed to be a random variable with distribution
\be
\langle V(\rv)\rangle =0, \ \langle V(\rv)V(\rv')\rangle=n_{i}v_0^2\delta (\rv -\rv').
\label{eq_37}
\ee
In the above $n_i$ is the impurity density and $v_0$ is the scattering amplitude.
Higher momenta can be present, and indeed they will be needed when considering skew-scattering processes, but it is not necessary to specify them for the time being.
Disorder effects can be taken into account in perturbation theory via  the inclusion of a self-energy.
\Eref{eq_3} becomes 
\be
\left[  \check G_0^{-1}(x_1,x_3)\overset{\otimes}{,} \check G(x_3,x_2) \right]=
\left[  \check \Sigma (x_1,x_3)\overset{\otimes}{,} \check G(x_3,x_2) \right].
\label{eq_38}
\ee
The lowest order self-energy due to disorder is given in  \fref{fig1} and its expression reads
\be
\check \Sigma_0 (p,x)=n_{i}v_0^2\sum_{\pv'} \check G(p,x).
\label{eq_39}
\ee
Notice that the integration is only on the space component of the $d$-momentum $p^\mu=(\ep, \pv')$. This is a result of the fact that the scattering is elastic. In order to use the above self-energy, we must transform it to the locally covariant formalism according to the transformation of  \eref{eq_18} and express $\check G$ in terms of $\check{\tilde G}$ via \eref{eq_19}. This procedure is the same we have followed in the previous section to transform the kinetic equation from the form of \eref{eq_11} to the form \eref{eq_27}. Since the procedure will also be used  later on, let us show it in detail in this simple case.
First we notice that 
\be
U_{\Gamma}(x,x_1) \left[ \check\Sigma (x_1,x_3)\overset{\otimes}{,}\check G(x_3,x_2)\right] U_{\Gamma}(x_2,x)=
\left[\check{\tilde \Sigma} (x_1,x_3)\overset{\otimes}{,}\check{\tilde G}(x_3,x_2) \right]
\label{eq_40}
\ee
after using the unitarity of the Wilson line by inserting 
$$U_{\Gamma}(x_3,x) U_{\Gamma}(x,x_3)=1$$
 between the self-energy and the Green function. The locally covariant self-energy reads 
\be 
\check{\tilde \Sigma}_0 =n_{i}v_0^2\sum_{\pv'}\left( \check{\tilde G}_{\pv'}+\frac{1}{2}\Big\{A^{\mu}(\partial_{p',\mu}-\partial_{p,\mu}),\check{\tilde G}_{\pv'}\Big\}\right)=n_{i}v_0^2\sum_{\pv'} \check{\tilde G}_{\pv'}.
\label{eq_41}
\ee
In the above the derivative with respect to $\ep$ cancels  in the two terms. The derivative with respect to $\pv$ vanishes because there is no dependence on $\pv$. Finally, the derivative
with respect to $\pv'$ can be integrated giving at most a constant, which can be discarded.
As a result, the locally covariant self-energy has the same functional form of the original self-energy. 
The Keldysh component of the collision integral then reads
\be
\tilde I=-i\left[ \check{\tilde \Sigma}, \check{\tilde G}_{\pv}\right]^K=
-in_iv_0^2\sum_{\pv'}\left( (\tilde G^R_{\pv'}-\tilde G^A_{\pv'})\tilde G^K_{\pv}-
(\tilde G^R_{\pv}-\tilde G^A_{\pv})\tilde G^K_{\pv'}\right).
\label{eq_42}
\ee
Note that the SU(2) shifted retarded and advanced Green functions have no spin structure and therefore commute with the Keldysh Green function.
For weak scattering, one can ignore the broadening of the energy levels in the retarded and advanced Green function and use
\be
\tilde G^R_{\pv}-\tilde G^A_{\pv}=-2\pi i \delta (\ep -\ep_{\pv}), \
\tilde G^K_{\pv}=-2\pi i \delta (\ep -\ep_{\pv}) (1-2f(\pv,\rv,t)).
\label{eq_43}
\ee
As a result, \eref{eq_34} is no longer collisionless and becomes
\be
\left(\tilde\partial_t +\frac{\pv}{m}\cdot \tilde \nabla_{\rv}\right)f(\pv, \rv, t)
-
\frac{e}{2}\Big\{ \left( {\bf E}+\frac{\pv}{m}\times {\bf B}\right)\cdot\nabla_{\pv},f(\pv, \rv, t)\Big\}=I[f],
\label{eq_44}
\ee
with the collision integral being
\be
I[f]=-2\pi n_i v_0^2\sum_{\pv'} \delta (\ep_{\pv}-\ep_{\pv'}) (f(\pv,\rv,t)-f(\pv',\rv,t)).
\label{eq_45}
\ee

It is appropriate in the final part of this section to obtain the solution of the Boltzmann equation \eref{eq_44}  in the diffusive approximation. First we notice that, by integration over the momentum $\pv$, the collision integral $I$ vanishes reproducing the continuity equation (\ref{eq_36}) with density and current defined in \eref{eq_35}. In the diffusive approximation we expand the distribution function in spherical harmonics
\be
f(\pv,\rv,t)=\langle f\rangle +2 {\hat \pv}\cdot {\bf f}+\dots
\label{eq_46}
\ee
and keep terms up to the p-wave symmetry. In the above $\langle\dots\rangle$ indicates the integration over the directions of the momentum. We perform the evaluation in two space dimensions having in mind the application of the theory to the 2DEG.
By defining the momentum relaxation time\footnote{Its expression can also be obtained, for instance, by the Fermi golden rule.}
\be
\frac{1}{\tau}=2\pi N_0 n_i v_0^2
\label{eq_47}
\ee
with the density of states $N_0=m/(2\pi)$ the collision integral becomes
\be
I[f]=-\frac{1}{\tau} 2{\hat \pv}\cdot {\bf f}.
\label{eq_48}
\ee
In the diffusive approximation we consider $\omega\tau \ll 1$ and $v_F  q\tau\ll 1$, where
$\omega$ and $q$ are typical energy and momentum scales. For instance, $\omega$ can be the magnitude of an externally applied magnetic field. We multiply \eref{eq_44} by ${\hat \pv}$ and integrate over the angle $\phi$ with ${\hat \pv}= (\cos\phi, \sin\phi)$. We get
\be
-\frac{1}{\tau}{\bf f}=\frac{p}{2m}\tilde \nabla_{\rv} \langle f\rangle-
\frac{e}{2}\langle \Big\{{\hat\pv}  {\bf E}\cdot \nabla_{\pv}, \langle f\rangle\Big\}\rangle-
\frac{e}{2m}\langle \Big\{{\hat\pv}  (\pv \times {\bf B}\cdot \nabla_{\pv}), 2 {\hat \pv}\cdot {\bf f} \Big\}\rangle.
\label{eq_49}
\ee
The first term, keeping in mind \eref{eq_35} for the current,  represents the {\it diffusive} contribution including the additional part  due to the SU(2) gauge field. Due to the covariant nature of the derivative, such a term differs from zero even in uniform circumstances. 
The second term yields the usual {\it drift} contribution, whereas the  third one gives rise to a Hall contribution.
The gradient with respect to the momentum can be split as $\nabla_{\pv}={\hat\pv}\partial_p -{\hat {\boldsymbol \phi}}\partial_{\phi}/p$ where ${\hat {\boldsymbol \phi}}= (-\sin\phi, \cos\phi )$. Then we get 
\be
{\bf f}=-\frac{\tau p}{2m}\tilde \nabla_{\rv} \langle f\rangle+\frac{e\tau }{4}\Big\{{\bf E},\partial_p \langle f\rangle \Big\}+\frac{e\tau}{2m}\Big\{ {\bf B}\times ,  {\bf f}\Big\}.
\label{eq_50}
\ee
By using the definitions of density and current in \eref{eq_36}, we may write the expression for the number and spin components as
\be
n={\rm Tr} \left[\rho\right], \ {\bf J}^0={\rm Tr}\left[{\bf J}\right], \ s^a=\frac{1}{2}{\rm Tr}\left[\sigma^a \rho\right], 
\ {\bf J}^a=\frac{1}{2}{\rm Tr}\left[\sigma^a {\bf J}\right].
\label{eq_51}
\ee
To begin with, let us consider the drift term 
\be
{\bf J}_{dr}=\sum_{\pv} \frac{p}{m}\frac{e\tau }{4}\Big\{{\bf E},\partial_p \langle f\rangle \Big\}=eN_0 \int {\rm d}\ep_\bp \ D(\ep_\bp) \frac{1}{2}\Big\{\partial_{\ep_\bp} \langle f\rangle , {\bf E}\Big\}=-\frac{e}{2}\Big\{ \sigma (\mu), {\bf E}\Big\}
\label{eq_52}
\ee
where $\ep_\bp = p^2/2m$, $D(\ep_\bp)= \tau \ep_\bp /m$, $\mu=\rho/N_0$ is a spin-dependent chemical potential, and $\sigma (\mu)=N_0 D(\mu)$. In equilibrium, $\rho_{eq}=N_0\ep_F+N_0\Phi$ and its eigenvalues determine the chemical potentials for {\it up} and {\it down} electrons. Then
\be
{\bf J}_{dr}^0=-eN_0 D^0{\bf E}^0 -\frac{e}{2}N_0 D^a{\bf E}^a, \
{\bf J}_{dr}^a=-\frac{e}{4}N_0 D^0{\bf E}^a -\frac{e}{2}N_0 D^a{\bf E}^0,
\label{eq_53}
\ee
with $D^0$ and $D^a$ defined by $D(\mu) = D^0 + D^a\sigma^a$.
By expanding around $\ep_F$, one has $D^0 \approx D(\epsilon_F) $ and $D^a \approx \tau s^a/(N_0 m)$, and therefore
\be
\sigma (\mu)=N_0 D(\ep_F) \sigma^0 +\frac{\tau}{m}s^a\sigma^a.
\label{eq_54}
\ee

The diffusion term is obtained by integrating over the momentum the first term of \eref{eq_50}
\be
{\bf J}_{dif}=-N_0 \int {\rm d}\ep_\bp \ D(\ep_\bp) \tilde \nabla_{\rv} \langle f\rangle=
-\frac{1}{2}\Big\{ D(\mu), \tilde \nabla_{\rv}\rho\Big\}.
\label{eq_55}
\ee
The above form of the diffusion term is determined by requiring that in equilibrium it must cancel the drift term according to the Einstein argument.
Then
\be
{\bf J}_{dif}^0=-D^0\nabla_{\rv}n -2  D^a \left[\tilde \nabla_{\rv }s\right]^a,\  {\bf J}_{dif}^a=-\frac{1}{2}D^a\nabla_{\rv}n -D^0 \left[\tilde \nabla_{\rv }s\right]^a.
\label{eq_56}
\ee
Finally the Hall term yields
\be
{\bf J}^0_{Hall}=\frac{e\tau}{m}{\bf B}^0\times {\bf J}^0+\frac{e\tau}{m}{\bf B}^a\times {\bf J}^a, \ \
{\bf J}^a_{Hall}=\frac{e\tau}{m}{\bf B}^0\times {\bf J}^a+\frac{e\tau}{4m}{\bf B}^a\times {\bf J}^0.
\label{eq_57}
\ee
To summarize, we may write the particle and spin currents as 
\ber
{\bf J}^0&=&-eN_0 D^0{\bf E}^0 -\frac{e}{2}N_0 D^a{\bf E}^a-D^0\nabla_{\rv}n -2  D^a \left[\tilde \nabla_{\rv }s\right]^a\nn\\
&+&
\frac{e\tau}{m}{\bf B}^0\times {\bf J}^0+\frac{e\tau}{m}{\bf B}^a\times {\bf J}^a
\label{eq_58}
\eer
and
\ber
{\bf J}^a&=&-\frac{e}{4}N_0 D^0{\bf E}^a -\frac{e}{2}N_0 D^a{\bf E}^0
-\frac{1}{2}D^a \nabla_{\rv}n -D^0 \left[\tilde \nabla_{\rv }s\right]^a\nn\\
&+&\frac{e\tau}{m}{\bf B}^0\times {\bf J}^a+\frac{e\tau}{4m}{\bf B}^a\times {\bf J}^0.
\label{eq_59}
\eer
The above two equations, together with the continuity-like \eref{eq_36},  will be used in the next section to analyze the spin Hall and Edelstein effect in the disordered Rashba model.

\section{The disordered Rashba model}\label{sec3a}
The Rashba Hamiltonian reads\cite{Rashba84}
\begin{equation}
\label{rash_1}
H=\frac{p^2}{2m}+\alpha p_y\sigma^x-\alpha p_x\sigma^y.
\end{equation}
The only non zero components  of the SU(2) gauge field are
\begin{equation}
\label{rash_2}
 e A_x^y=-2m\alpha,  \
 e A_y^x=2m\alpha.  \
\end{equation}
As shown in the previous sections (cf. \eref{eq_36}), the spin density obeys a continuity-like equation
\begin{equation}
\label{rash_3}
\tilde\partial_t s^a +\tilde\nabla_{\rv}\cdot {\bf J}^a=0,
\end{equation}
which is deceptively simple. The notable fact in the present theory is that the covariant derivatives defined in \eref{eq_28} appear also at the level of the effective phenomenological equations, providing an elegant and compact way to derive the equation of motion for the spin density. 
The explicit expressions of the space and time covariant derivatives 
of   a generic observable ${\cal O}^a$  read
\footnote{$\epsilon_{abc}$ is the fully antisymmetric Ricci tensor.}
\begin{eqnarray}
\tilde\partial_t {\cal O}^a&=& \partial_t {\cal O}^a +\epsilon_{abc} \  e \Phi^b {\cal O}^c \label{covtime}\\
\tilde\nabla_{i}{\cal O}^a &=&\nabla_i {\cal O}^a -\epsilon_{abc} \ e  A_i^b {\cal O}^c \label{covspace}.
\end{eqnarray}
Equation (\ref{rash_3}) becomes
\begin{equation}
\label{continuityexp}
\partial_t s^a +\epsilon_{abc}  \ e  \Phi^b s^c+\nabla_i J_i^a-\epsilon_{abc} \  e  A_i^b J^c_i=0,
\end{equation}
showing that the equation for the spin is not a simple continuity equation, as expected from the non conservation of spin. The second term in Eq.~(\ref{continuityexp}) is the standard {\it precession} term. The last term of (\ref{continuityexp}) can be made explicit by providing the expression for the spin current $J_i^a$, where the lower (upper) index indicates the space (spin) component. 
The expression of $J^a_i$ was derived via a microscopic theory in the diffusive regime in \eref{eq_59}. The explicit expression reads
\begin{equation}
\label{spincurrent}
J_i^a=-\frac{e\tau}{m}s^aE_i +D\epsilon_{abc}  e   A_i^b s^c-\frac{ e  \tau }{4m}\epsilon_{ijk}J_j \ B^a_k -\frac{e  DN_0}{2}  E^a_i ,
\end{equation}
where $D=D(\ep_F)$ is the diffusion constant.
Let us apply Eqs.~(\ref{continuityexp}-\ref{spincurrent}) to the Rashba model defined by \eref{rash_2} in the presence of an applied electric field along the x direction $E_x$. To linear order in the electric field, the first term of \eref{spincurrent} does not contribute  in a paramagnetic system. Also by using \eref{rash_2} in the expressions of the fields of \eref{eq_32} we get that the  SU(2) electric field vanishes  $E^a_i=0$ and that the only non zero component of the SU(2) magnetic field reads
\begin{equation}
\label{spinhallfield}
e  B^z_z= -(2m\alpha )^2.
\end{equation}
Because the electric field is uniform, we may ignore the space derivative and
obtain the explicit form of \eref{continuityexp}
\begin{eqnarray}
\partial_t s^x&=&-2m\alpha J^z_x  \label{conx}\\
\partial_t s^y&=& -2m \alpha J^z_y  \label{cony}\\
\partial_t s^z&=&
 +2m \alpha J^y_y
+2m\alpha J^x_x 
  \label{conz}
\end{eqnarray}
with the associated  expressions for the spin currents
\begin{eqnarray}
J^z_x&=&2m\alpha D s^x 
\label{currzx}\\
J^z_y&=& 2 m\alpha D s^y +\theta^{int}_{SH} J_x^0
 \label{currzy}\\
J^x_x=J^y_y&=&  -2m\alpha D s^z, \label{curryy}
\end{eqnarray}
where $J_x^0=-e \sigma(\ep_F) E_x$ is the charge current and $\theta_{SH}^{int}$ is  the spin Hall angle for  intrinsic SOC
\begin{equation}
\label{shaint}
\theta_{SH}^{int} =-m\tau \alpha^2.
\end{equation}
Insertion of Eqs.~(\ref{currzx}-\ref{curryy}) into (\ref{conx}-\ref{conz}) gives the Bloch equations 
\ber
\partial_t s^x&=&-\frac{1}{\tau_{DP}}s^x\label{eq_69a}\\
\partial_t s^y &=&-\frac{1}{\tau_{DP}}(s^y -s_0)
\label{eq_69}\\
\partial_t s^z&=&-\frac{2}{\tau_{DP}}s^x\label{eq_69b}
\eer
where the Dyakonov-Perel relaxation time is given by 
\be
\tau_{DP}^{-1}=(2m\alpha)^2 D
\label{eq_69c}
\ee
and the current-induced spin polarization is given by
\be
s_0=-eN_0\alpha\tau E_x.
\label{edelstein}
\ee
In the static limit the solution of the Bloch equations yields an in-plane spin polarization perpendicular to the electric field $s^y=s^0$, with $s^x=s^z=0$. This is known as the Edelstein\cite{Edelstein90,Aronov89} or inverse spin-galvanic effect\cite{Ganichev02,Ganichev2016}. 
The vanishing of the time derivative implies, via \eref{cony}, the vanishing of the spin current $J^z_y$ associated to the spin Hall effect. This vanishing occurs thanks to the exact compensation of the two contributions appearing in \eref{currzy}.
\footnote{To make contact with the diagrammatic Kubo formula approach, we notice that the second term of \eref{currzy} corresponds to a bubble-like diagram, whereas the first term describes the so-called vertex corrections.\cite{Raimondi05}}

The above analysis can be extended in the presence of a thermal gradient.
More precisely, we derive $s^y$ in terms of a stationary thermal gradient along the  $x$-direction, $\nabla_x T$, in the absence of any additional external fields. \cite{toelle2014} 

However, in an experiment one would still measure an electric field $E_x$ due to a gradient in the chemical potential, $\nabla_x \mu$, resulting from an imbalance of the charge carriers due to the thermal gradient. For this, we shall first consider the trace of \eref{eq_50}:
\be \label{TE_bff}
{\bf f}^0 = -\frac{\tau p}{2m} \nabla \langle f^0 \rangle  ,
\ee
where we can approximate $\nabla \langle f^0 \rangle $ as the Fermi function $f^{eq}$, giving us
\be \label{TE_fx0}
f_x^0  = - \frac{\tau p}{2m} \left(\frac{\epsilon - \mu}{T} \nabla_x T + \nabla_x \mu \right) \left( - \frac{\partial f^{eq}}{\partial \epsilon} \right)
\ee
for the $x$-component of \eref{TE_bff}. With use of the Sommerfeld expansion
\be \label{TE_Sommer}
\int d\epsilon g(\epsilon) \left(- \frac{\partial f^{eq}}{\partial \epsilon} \right) = g(\mu) + \frac{\pi^2}{6} \left( k_B T \right)^2 \left.\frac{\partial^2 g}{\partial \epsilon^2}\right|_{\epsilon = \mu}  ,
\ee
where $g(\epsilon)$ is an arbitrary energy dependent function, we end up with the particle current in the $x$-direction as follows:
\be
J_x^0 = -\frac{2\tau N_0}{m} \left[\frac{\pi^2}{3} \frac{(k_B T)^2}{T} \nabla_x T + \mu \nabla_x \mu \right]  .
\ee
We shall consider an open circuit along $x$-direction, i.e., a vanishing particle current $J_x^0 = 0$. Then, we can express the electric field one would measure in an experiment as
\be \label{TE_Seebeck}
E_x = \frac{1}{e} \nabla_x \mu = S \nabla_x T ,
\ee
where $S = - (\pi k_B)^2 T / (3e\mu)$  is the Seebeck coefficient.
After having analyzed the charge sector,  we  consider next the spin sector in order to get an expression for $s^y$.  We start by multiplying \eref{eq_50} with $\sigma^z$ and perform the trace. The $y$-component reads
\be \label{TE_fyz}
f_y^z = 2p\tau\alpha \langle f^y \rangle + \frac{ e  \tau}{m} B_z^z f_x^0 \, .
\ee
Note that the form of \eref{cony} doesn't change and since we assume a stationary case we have
\be 
J_y^z = 0 \Leftrightarrow f_y^z = 0 \, .
\ee
This implies that there is no  spin Nernst effect as no spin Hall effect in the disordered Rashba model.
From \eref{TE_fyz}, together with \eref{TE_fx0} we therefore find
\be
\langle f^y \rangle = \frac{2\alpha m}{p} f_x^0 = - \alpha\tau \left(\frac{\epsilon - \mu}{T} \nabla_x T + \nabla_x \mu \right) \left( - \frac{\partial f^{eq}}{\partial \epsilon} \right) \, .
\ee
From the form of the latter equation and with use of the Sommerfeld expansion, \eref{TE_Sommer}, it is clear that we can express the $y$ spin polarization as
\be
s^y = P_{sT} \nabla_x T + P_{sE} E_x \, ,
\ee
where $P_{sT}$ can be written in a Mott-like form:
\be 
P_{sT} = - S \mu \frac{\partial P_{sE}}{\partial \mu} .
\ee
Here, we find $P_{sE} = -\alpha \tau e N_0 $, consistent with \eref{eq_69}.  This results in a vanishing $P_{sT}$ since $P_{sE}$ is independent of $\mu$. We express $E_x$ in terms of $\nabla_x T$ with use of \eref{TE_Seebeck} and end up with 
\be 
s^y = -\alpha\tau e N_0 S \nabla_x T ,
\ee
describing the thermal Edelstein effect in the disordered Rashba model.

\section{The impurity-induced spin-orbit coupling: swapping and side jump mechanisms}\label{sec4}

In this and following sections we consider the {\it extrinsic} SOC due to impurity scattering described by the Hamiltonian
\be
H_{ext, so}=-\frac{\lambda_0^2}{4}{\boldsymbol \sigma} \times \nabla V(\rv )\cdot \pv,
\label{eq_70}
\ee
where $\lambda_0$ is the effective Compton wave length\cite{Engel05,Tse06}.
In developing the perturbation theory in the impurity potential we must now use the lowest order scattering amplitude
\be
\label{eq_71}
S_{\pv',\pv''}=V_{\pv'-\pv''}\left[ 1-\frac{i\lambda_0^2}{4} \pv'\times 	\pv'' \cdot {\boldsymbol \sigma}\right]
\ee
with
\be
\langle V_{\qv_1}V_{\qv_2}\rangle=n_i v_0^2\delta (\qv_1+\qv_2).
\label{eq_72}
\ee
\begin{figure}
\begin{center}
\includegraphics[width=3.in]{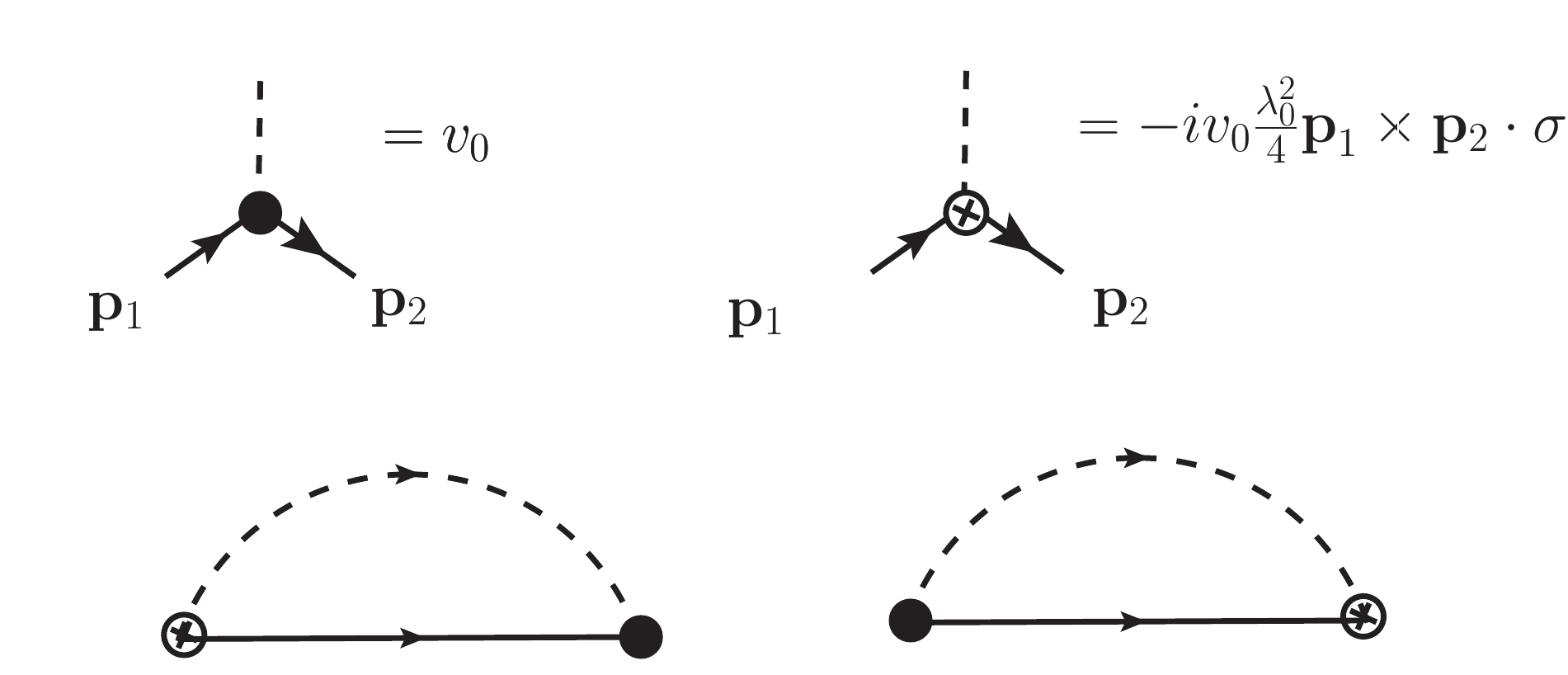}
\caption{In the top line the impurity insertion without (black dot vertex) and with (crossed dot vertex) spin-orbit coupling. In the bottom line the two diagrams to first order in the spin-orbit coupling $\lambda_0^2$.}
\label{fig2}
\end{center}
\end{figure}
To zeroth order  in $\lambda_0^2$, we have the diagram  of \fref{fig1}, which has been studied in the previous section. To first order in $\lambda_0^2$, we must consider the two diagrams  of  \fref{fig2}. Here an empty crossed dot stands for the part of the scattering amplitude with the spin-orbit coupling. Let us evaluate these diagrams step by step. Before the impurity average (indicated as $\langle \dots \rangle$), the expression for the two diagrams reads
\ber
\check \Sigma_{1, \pv',\pv''}&=&-i\frac{\lambda_0^2}{4}\sum_{\pv_1,\pv_2}\langle V_{\pv'-\pv_1}\left(
\pv'\times \pv_1 \cdot {\boldsymbol \sigma} \check G_{\pv_1,\pv_2}\right.\nn\\
&+&\left.
\check G_{\pv_1,\pv_2}\pv_2\times \pv'' \cdot {\boldsymbol \sigma} \right)  V_{\pv_2-\pv''} \rangle.
\label{eq_73}
\eer
In the above, $\check G_{\pv_1,\pv_2}$ is the Fourier transform with respect to the space arguments $\rv_1$ and $\rv_2$ of $\check G(\rv_1,\rv_2)$. We do not mention explicitly here the time arguments for the sake of simplicity. Performing the impurity average one obtains
$\pv'-\pv_1=\pv''-\pv_2$. It is convenient then to define momenta as 
\be
\pv =\frac{\pv'+\pv''}{2}, \ \qv=\pv'-\pv''=\pv_1-\pv_2, \  {\tilde \pv}=\frac{\pv_1+\pv_2}{2}
\label{eq_74}
\ee
in such a way that $\pv$ and ${\tilde \pv}$ correspond to the momentum of the mixed Wigner representation introduced previously. The momentum $\qv$ instead is the variable conjugated to the center-of-mass coordinate $\rv$  by Fourier transform. We then get the impurity-averaged   expression of the two first-order diagrams
\ber
\check \Sigma_1 (\pv,\qv)&=&-i\frac{\lambda_0^2}{4}n_iv_0^2\sum_{\tilde{\pv}}
\left[   \left(\pv+\frac{\qv}{2}\right)\times \left( \tilde{\pv}+\frac{\qv}{2}\right) \cdot {\boldsymbol \sigma} \check G(\tilde{\pv},\qv)\right.\nn\\
&+&\left. \check G(\tilde{\pv},\qv)\left(\tilde{\pv}-\frac{\qv}{2}\right)\times \left(\pv-\frac{\qv}{2}\right) \cdot {\boldsymbol \sigma}
\right].
\label{eq_75}
\eer
The above expression can be divided into three terms
\begin{eqnarray}
\check \Sigma_{1,a} (\pv,\qv))& = &- i\frac{\lambda_0^2}{4}n_iv_0^2\sum_{\tilde{\pv}}\left[ \pv\times  \tilde{\pv} \cdot {\boldsymbol \sigma},  \check G(\tilde{\pv},\qv)\right] \label{eq_76}\\
\check \Sigma_{1,b} (\pv,\qv)) & = &-i\frac{\lambda_0^2}{8}n_iv_0^2\sum_{\tilde{\pv}} \Big\{ \pv\times \qv \cdot {\boldsymbol \sigma}, \check  G(\tilde{\pv},\qv)\Big\} \label{eq_77}\\
\check\Sigma_{1,c}  (\pv,\qv))&=&i\frac{\lambda_0^2}{8}n_iv_0^2\sum_{\tilde{\pv}}  \Big\{ \tilde{\pv}\times\qv \cdot {\boldsymbol \sigma}, \check  G(\tilde{\pv},\qv)\Big\} .\label{eq_78}
\end{eqnarray}
One can Fourier transform back to the center-of-mass coordinate $\rv$ and re-label $\tilde\pv \rightarrow \pv'$
\begin{eqnarray}
\check \Sigma_{1,a}  (\pv,\rv))& = &- i\frac{\lambda_0^2}{4}n_iv_0^2\sum_{{\pv'}}\left[ \pv\times  {\pv'} \cdot {\boldsymbol \sigma},  \check G({\pv'},\rv)\right] \label{eq_79}\\
\check \Sigma_{1,b} (\pv,\rv)) & = &-\nabla_{\rv} \cdot \frac{\lambda_0^2}{8}n_iv_0^2\sum_{{\pv'}} \Big\{{\boldsymbol \sigma}\times  \pv    , \check  G({\pv'},\rv)\Big\} \label{eq_80}\\
\check\Sigma_{1,c}  (\pv,\rv))&=&\nabla_{\rv}\cdot  \frac{\lambda_0^2}{8}n_iv_0^2\sum_{{\pv'}}  \Big\{ {\boldsymbol \sigma} \times  \pv'  , \check  G({\pv'},\rv)\Big\} .\label{eq_81}
\end{eqnarray}
Equations (\ref{eq_79}-\ref{eq_81}) are the final expression for the diagrams of the second line of \fref{fig2}. The first term $\check \Sigma_{1,a}$, as we will see, describes the swapping of spin currents under scattering\cite{Lifshitz2009}. The other two terms,
$\check \Sigma_{1,b}$ and $\check\Sigma_{1,c}$, are written as a divergence. As a consequence, when they are inserted in the collision integral of the kinetic equation, they lead to a correction of the velocity operator and hence describe the so-called {\it side-jump} mechanism\cite{Berger1970}.

Until now we have not yet considered the effect of the gauge fields on the extrinsic SOC. To do it, we must transform the above derived self-energy to the locally covariant form. 
Let us first examine the U(1) gauge field corresponding to a static electric field
${\bf E}^0=-\nabla_{\rv }\Phi^0(\rv)$.  The  shift of the gradient of the Green function yields
\ber
\widetilde{\nabla_{\rv} \check G(\pv,\rv)}&=&\nabla_{\rv} \check G(\pv,\rv)-\frac{1}{2}
\Big\{e  \Phi^0\partial_{\ep},\nabla_{\rv} \check G(\pv,\rv)\Big\}\nn \\
&=&\nabla_{\rv} \check{\tilde G}(\pv,\rv)+\frac{1}{2}\nabla_{\rv}
\Big\{ e  \Phi^0\partial_{\ep}, \check{\tilde G}(\pv,\rv)\Big\}
-\frac{1}{2}
\Big\{ e  \Phi^0\partial_{\ep},\nabla_{\rv} \check{\tilde G}(\pv,\rv)\Big\}\nn \\
&=&\nabla_{\rv} \check{\tilde G}(\pv,\rv)-e  {\bf E}^0 \partial_{\ep} \check{\tilde G}(\pv,\rv).
\label{eq_82}
\eer
As a result we get from Eqs.(\ref{eq_80}-\ref{eq_81}) two more terms
\begin{eqnarray}
\delta \check \Sigma_{1,b} (\pv,\rv)) & = & e  {\bf E}^0 \cdot \frac{\lambda_0^2}{8}n_iv_0^2\sum_{{\pv'}} \Big\{{\boldsymbol \sigma}\times  \pv    , \partial_{\ep}\check  {\tilde G}({\pv'},\rv)\Big\} \label{eq_83}\\
\delta \check\Sigma_{1,c}  (\pv,\rv))&=&- e  {\bf E}^0 \cdot  \frac{\lambda_0^2}{8}n_iv_0^2\sum_{{\pv'}}  \Big\{ {\boldsymbol \sigma} \times  \pv'  , \partial_{\ep}\check  {\tilde G}({\pv'},\rv)\Big\} .\label{eq_84}
\end{eqnarray}
Let us transform \eref{eq_79} to the locally covariant form
\ber
\check{\tilde \Sigma}_{1,a}(\pv,\rv)&=&- i\frac{\lambda_0^2}{4}n_iv_0^2\sum_{{\pv'}}\left[ \pv\times  {\pv'} \cdot {\boldsymbol \sigma},  \check{\tilde G}({\pv'},\rv)\right]  \nn\\
&+& i\frac{\lambda_0^2}{4}n_iv_0^2\sum_{{\pv'}} \frac{1}{2}\left[ \pv\times  {\pv'} \cdot {\boldsymbol \sigma},  \Big\{ e  {\bf A}\cdot \nabla_{\pv'}, \check{\tilde G}({\pv'},\rv)\Big\}\right]  \nn \\
&-&  i\frac{\lambda_0^2}{4}n_iv_0^2\sum_{{\pv'}}
\frac{1}{2}\Big\{ e  {\bf A}\cdot \nabla_{\pv}, 
\left[ \pv\times  {\pv'} \cdot {\boldsymbol \sigma},  \check{\tilde G}({\pv'},\rv)\right] \Big\},\nn
\eer
which can be rewritten as
\ber
\check{\tilde \Sigma}_{1,a}(\pv,\rv)&=&- i\frac{\lambda_0^2}{4}n_iv_0^2\sum_{{\pv'}}\left[ \pv\times  {\pv'} \cdot {\boldsymbol \sigma},  \check{\tilde G}({\pv'},\rv)\right]  \nn\\
&+& i\frac{\lambda_0^2}{4}n_iv_0^2\sum_{{\pv'}} \frac{1}{2}\left[ e  {\bf A}\overset{\cdot}{  ,}  \Big\{ {\boldsymbol \sigma} \times  {\pv}  , \check{\tilde G}({\pv'},\rv)\Big\}\right]  \nn \\
&-&  i\frac{\lambda_0^2}{4}n_iv_0^2\sum_{{\pv'}}
\frac{1}{2}\Big\{{\boldsymbol \sigma} \times  {\pv'} \overset{\cdot }{, }
\left[   { e  \bf A},  \check{\tilde G}({\pv'},\rv)\right] \Big\}.
\label{eq_85}
\eer
As a result, the contribution of the diagrams of  \fref{fig2} can be written
as
\ber
\check{\tilde \Sigma}_{1,a}^{SCS}&=&- i\frac{\lambda_0^2}{4}n_iv_0^2\sum_{{\pv'}}\left[ \pv\times  {\pv'} \cdot {\boldsymbol \sigma},  \check{\tilde G}({\pv'},\rv)\right] \label{eq_86}\\
\check{\tilde \Sigma}_{1,b}^{SJ}\ &=&-\frac{\lambda_0^2}{8}n_iv_0^2\sum_{\pv'} \tilde \nabla_{\rv} \Big\{{\boldsymbol \sigma }\times \pv ,\check{\tilde G}(\pv',\rv)\Big\}\label{eq_87}\\
\check{\tilde \Sigma}_{1,c}^{SJ}\ &=&\ \frac{\lambda_0^2}{8}n_iv_0^2\sum_{\pv'}\Big\{ {\boldsymbol\sigma}\times \pv' , \tilde \nabla_{\rv}\check{\tilde G}(\pv',\rv)\Big\}. \label{eq_88}
\eer
where $\tilde \nabla_\rv=\nabla_{\rv}-e {\bf E}^0\partial_{\ep} +i \left[ e {\bf A}, \dots\right]$. 
A few comments are needed at this point. As already mentioned, the term $\check{\tilde \Sigma}_{1,a}^{SCS}$ describes the spin current swapping (SCS) defined by
\be
J^a_i=\kappa \left[ J^i_a-\delta_{ia}\sum_l J^l_l\right].
\label{scs_def}
\ee
 This is evident in the fact that this term contains the vector product $\pv\times \pv'$ of the momenta before and after the scattering from the impurity. The presence of both momenta yields the coupling of the currents of incoming and outgoing particles. The other two terms $\check{\tilde \Sigma}_{1,b}^{SJ}$ and $\check{\tilde \Sigma}_{1,c}^{SJ}$ describe the so-called side jump (SJ) effect. This is evident in both terms, which show the operator ${\boldsymbol \sigma}\times (\pv -\pv')$. By a semiclassical analysis one can show that $\Delta \rv \equiv -(\lambda_0^2/4)(\pv' -\pv)\times {\boldsymbol\sigma}$ is the side jump shift caused by the SOC to the scattering trajectory of a wave packet. The SJ terms are composed of three parts. The first part is the one under the space derivative sign and can be written as $-\nabla_{\rv}\cdot \delta {\bf J}^{SJ}$, i.e. it describes a modification of the current operator. Eventually  this term yields the first {\it one-half} of the side jump. The second part, proportional to the electric field, describes how the energy of a scattering particle is affected by the effective dipole energy $\sim  e  {\bf E}^0\cdot \Delta \rv$ due to the side jump shift of the trajectory. This is the other {\it one-half} contribution to the side jump. The third part reconstructs the full covariant derivative in the presence of a SU(2) gauge field ${\bf A}={\bf A}^a\sigma^a/2$. To first order accuracy  in the gradient expansion, we have replaced the $\check G$ with $\check{\tilde G}$ in the terms where the gradient or the gauge field appear.
To make the above comments explicit, we start by noticing that only the Keldysh component appears in \eref{eq_86} because both $\tilde G^R$ and $\tilde G^A$ are proportional to $\sigma^0$ and commute with all the Pauli matrices. By considering that the Keldysh component of the collision integral requires $\tilde \Sigma^R \tilde G^K -\tilde G^K\tilde \Sigma^A- (\tilde G^R-\tilde G^A) \tilde \Sigma^K$, one obtains\cite{Shenprb14} from \eref{eq_86}
\be
I^{SCS}[f]=-i \frac{\lambda_0^2}{4}n_i v_0^2 \sum_{\pv'} \delta (\ep_{\pv}-\ep_{\pv'})\left[ {\boldsymbol\sigma}\cdot \pv\times \pv',f_{\pv'}\right].
\label{eq_89}
\ee
Similarly,  for the side jump we define 
\be
I^{SJ}=-i\int \frac{{\rm d}\ep}{4\pi i} \left(- (\tilde G^R-\tilde G^A) \tilde \Sigma^K+\tilde \Sigma^R \tilde G^K -\tilde G^K\tilde \Sigma^A \right)\equiv I^{(a)}+I^{(b)}.
\label{eq_90}
\ee
Because of the integration over the angle, the retarded and advanced components of \eref{eq_88} vanish. The retarded component of \eref{eq_87} reads
\be
\tilde \Sigma^{SJ, R}_{1,b}=-i 2\pi \frac{\lambda_0^2}{8}n_iv_0^2\sum_{\pv'}\delta (\ep_{\pv}-\ep_{\pv'})\tilde\nabla_{\rv} {\boldsymbol\sigma}\times \pv
\label{eq_91}
\ee
and $\tilde \Sigma^{SJ, A}_{1,b}=-\tilde \Sigma^{SJ, R}_{1,b}$. As a result,
with $h_{\pv}\equiv 1-2f_{\pv}$ for brevity,
\be
I^{(b)}=-\frac{\lambda_0^2}{16}2\pi n_iv_0^2 \sum_{\pv'}  \delta (\ep_{\pv}-\ep_{\pv'}) \Big\{\tilde\nabla_{\rv}( {\boldsymbol\sigma}\times \pv), h_{\pv}\Big\}.
\label{eq_92}
\ee
By using again the identity $\{A,\left[B,C\right]\}-\left[B,\{A,C\}\right]=\{\left[A,B\right],C\}$, the Keldysh component of \eref{eq_88} reads
\be
\tilde\Sigma^{SJ, K}_{1,c}=\frac{\lambda_0^2}{8}n_i v_0^2 \sum_{\pv'}
\left( \tilde\nabla_{\rv}\Big\{ {\boldsymbol \sigma}\times {\pv'}, \tilde G_{\pv'}^K\Big\}-
\Big\{ \tilde\nabla_{\rv}({\boldsymbol\sigma}\times \pv' ),\tilde G_{\pv'}^K\Big\}\right).
\label{eq_93}
\ee
By combining the last result with \eref{eq_87} and \eref{eq_92}, one obtains finally
\ber
I^{SJ}[f]&=& -\tilde \nabla_{\rv}\cdot \frac{\lambda_0^2}{8}n_i v_0^2 \sum_{\pv'}\delta (\ep_{\pv}-\ep_{\pv'}) \Big\{ {\boldsymbol\sigma}\times ( \pv'-\pv), f_{\pv'}\Big\}\label{eq_94}\\
&+&\frac{\lambda_0^2}{8}n_i v_0^2 \sum_{\pv'}\delta (\ep_{\pv}-\ep_{\pv'}) 
\left(\Big\{\tilde\nabla_{\rv}( {\boldsymbol\sigma}\times \pv'), f_{\pv'}\Big\}-\Big\{\tilde\nabla_{\rv}( {\boldsymbol\sigma}\times \pv), f_{\pv}\Big\} \right).
\nn
\eer
The first term on the right hand side, under the covariant space derivatiive, defines the modification of the current operator 
due to the SOC. We emphasize that such anomalous part of the current is subject to the full covariant derivative. Hence the last term in \eref{continuityexp} remains the same.
Notice also that the second term in $I^{SJ}[f]$, although it does not contribute to the continuity equation, is necessary to make sure that the equilibrium distribution function solves the kinetic equation.

\section{The impurity-induced spin-orbit coupling: skew scattering}\label{sec5}

\begin{figure}
\begin{center}
\includegraphics[width=3.in]{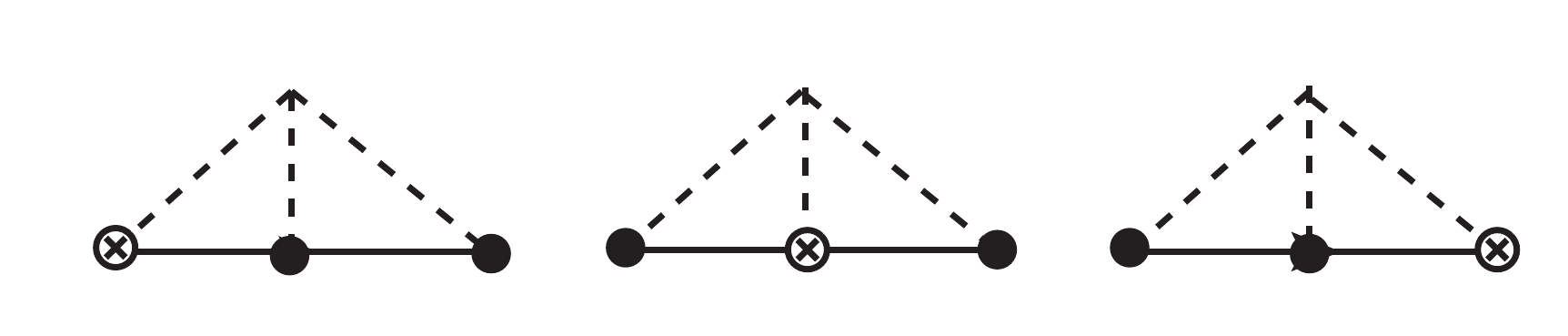}
\caption{Third order in $v_0$ and first order in $\lambda_0^2$ diagrams. The skew scattering contribution arises from the first and last diagram.}
\label{fig3}
\end{center}
\end{figure}

 In this section we  discuss skew scattering by considering the diagrams of  \fref{fig3}.
 To understand the meaning of these diagrams, recall that in general, in the presence of SOC, the scattering amplitude reads
\begin{equation}
\label{ss1}
S=A+{\hat {\bf p}}\times {\hat {\bf p}'}\cdot {\boldsymbol \sigma} B,
\end{equation}
where ${\hat {\bf p}}$ and ${\hat {\bf p}'}$ are unit vectors in the direction of the momentum before and after the scattering event. To lowest order in perturbation theory or Born approximation one has 
$A=v_0$ and $B=-{\rm i}(\lambda_0^2p_F^2 /4) v_0$ and one recovers \eref{eq_71}.
By considering the scattering probability proportional to $|S|^2$, one obtains three contributions given by $|A|^2$, $|B|^2$ and
$2 \ {\cal R}e \  (A B^*) {\hat {\bf p}}\times {\hat {\bf p}'}\cdot {\boldsymbol \sigma}$. 
Whereas the first two contributions are spin independent and give the total scattering time, 
the third one represents the so-called skew scattering term according to 
which electrons with opposite spin are scattered in opposite directions. 
Clearly, since $A$ and $B$ are out of phase, there is no skew scattering effect to the order of the Born approximation. 
For it to appear to first order in the spin-orbit coupling constant $\lambda_0^2$, 
$A$ has to be evaluated beyond the Born approximation. 
The scattering problem can be cast in terms of the Lippman-Schwinger equation
\begin{equation}
\label{ss2}
\psi ({\bf x}) =e^{{\rm i}{\bf k}\cdot {\bf x}}+\int {\rm d}{\bf x}' \ G ({\bf x}-{\bf x}')V({\bf x}')\psi ({\bf x}'),
\end{equation}
where $G({\bf x})$ is the retarded Green function at fixed energy.
From (\ref{ss2}) we get
\begin{equation}
\label{ss3}
\psi^{(1)}=v_0G({\bf x}), \ A^{(1)}=v_0; \ \ \psi^{(2)}=v_0^2 G({\bf 0}) G({\bf x}), \ A^{(2)}=v_0^2G({\bf 0}).
\end{equation}
Notice that only the imaginary part of $A^{(2)}$ is needed. By recalling that ${\cal I}m \  G({\bf 0})=-\pi N_0$, 
the spin-orbit independent scattering amplitude $A$ up to second order in $v_0$ reads
\begin{equation}
\label{ss4}
A=v_0 \left( 1-{\rm i}\pi N_0 v_0\right).  
\end{equation}
The skew scattering contribution will then follow by inserting the modified scattering amplitude (\ref{ss4})
into the collision integral of the Boltzmann equation.
The same result can, of course, be obtained in quantum field theory using the Green function technique. 
The latter becomes necessary when one wants to consider skew scattering 
in the presence of Rashba spin-orbit interaction. 
To this end one has to consider the electron self-energy at least to third order in the scattering potential $v_0$.  The diagrams of    \fref{fig3} yield 
\begin{eqnarray}
\check\Sigma_a^{SS} & = & -\frac{i\lambda_0^2}{4}\sum_{\pv_1,\pv_2,\pv_3,\pv_4}\langle V_{\pv'-\pv_1} \check G_{\pv_1,\pv_2}V_{\pv_2-\pv_3}\check G_{\pv_3,\pv_4}V_{\pv_4-\pv''}\pv_4\times \pv''\cdot{\boldsymbol \sigma} \rangle \nonumber\\
\check\Sigma_b^{SS} & = &- \frac{i\lambda_0^2}{4}\sum_{\pv_1,\pv_2,\pv_3,\pv_4} \langle V_{\pv'-\pv_1} \check G_{\pv_1,\pv_2}V_{\pv_2-\pv_3}\pv_2\times \pv_3\cdot{\boldsymbol \sigma} \check G_{\pv_3,\pv_4}V_{\pv_4-\pv''}\rangle \nonumber\\
\check\Sigma_c^{SS} &=&-\frac{i\lambda_0^2}{4}\sum_{\pv_1,\pv_2,\pv_3,\pv_4}\langle V_{\pv'-\pv_1} \pv'\times \pv_1\cdot {\boldsymbol \sigma} \check G_{\pv_1,\pv_2}V_{\pv_2-\pv_3} \check G_{\pv_3,\pv_4}V_{\pv_4-\pv''}\rangle . \nonumber
\end{eqnarray}
By requiring the existence of third moments of the random potential
$\langle V(\qv_1)V(\qv_2)V(\qv_3)\rangle =n_i v_0^3 \delta (\qv_1+\qv_2+\qv_3)$, we perform the impurity average and obtain
\ber
\check \Sigma_a^{SS} 
& = &-n_iv_0^3 \frac{i\lambda_0^2}{4}
\sum_{\pv_1,\pv_2,\pv_3,\pv_4}
\delta_{\pv'-\pv_1+\pv_2-\pv_3+\pv_4-\pv'' } \check G_{\pv_1,\pv_2} \check G_{\pv_3,\pv_4} \pv_4\times \pv''\cdot{\boldsymbol \sigma} 
 \nonumber\\
\check \Sigma_b^{SS} 
& = &-n_iv_0^3 \frac{i\lambda_0^2}{4}
\sum_{\pv_1,\pv_2,\pv_3,\pv_4}
\delta_{\pv'-\pv_1+\pv_2-\pv_3+\pv_4-\pv'' } \check G_{\pv_1,\pv_2} \pv_2\times \pv_3\cdot{\boldsymbol \sigma} \check G_{\pv_3,\pv_4}
 \nn\\
\check \Sigma_c^{SS} 
&=&-n_iv_0^3\frac{i\lambda_0^2}{4}
\sum_{\pv_1,\pv_2,\pv_3,\pv_4}
\delta_{ \pv'-\pv_1+\pv_2-\pv_3+\pv_4-\pv'' }
 \pv'\times \pv_1 \cdot {\boldsymbol \sigma} \check G_{\pv_1,\pv_2} \check G_{\pv_3,\pv_4} . 
 \nn
\eer
Let us introduce as before,  momenta associated to center-of-mass and relative coordinates
$
\pv  =  (\pv'+\pv'')/2, \ \qv  =  \pv'-\pv'' ,\ \widetilde{\pv}_a  =  (\pv_1+\pv_2)/2,
\widetilde{\qv}_a  =  \pv_1-\pv_2, \widetilde{\pv}_b  =  (\pv_3+\pv_4)/2, \widetilde{\qv}_b  =  \pv_3-\pv_4
$
and we get after  integrating over the momentum $\widetilde{\qv}_b$
\begin{eqnarray}
\check \Sigma_a^{SS} & = &-n_iv_0^3 \frac{i\lambda_0^2}{4}\sum_{\widetilde{\pv}_a, \widetilde{\pv}_b,\widetilde{\qv}_a} \check G_{\widetilde{\pv}_a,\widetilde{\qv}_a} \check G_{\widetilde{\pv}_b,\qv-\widetilde{\qv}_a}\left(\widetilde{\pv}_b-\frac{\qv}{2}+\frac{\widetilde{\qv}_a}{2}\right)\times \left(\pv-\frac{\qv}{2}\right)\cdot{\boldsymbol \sigma}  \nonumber\\
\check\Sigma_b^{SS} & = &-n_iv_0^3 \frac{i\lambda_0^2}{4}
\sum_{\widetilde{\pv}_a, \widetilde{\pv}_b,\widetilde{\qv}_a}
 \check G_{\widetilde{\pv}_a,\widetilde{\qv}_a} 
  \left( \widetilde{\pv}_a -\frac{\widetilde{\qv}_a}{2}\right) \times \left( \widetilde{\pv}_b+\frac{\qv}{2}-\frac{\widetilde{\qv}_a}{2}\right) 
  \cdot{\boldsymbol \sigma} \check G_{\widetilde{\pv}_b,\qv-\widetilde{\qv}_a} \nonumber\\
\check\Sigma_c^{SS} &=&-n_iv_0^3\frac{i\lambda_0^2}{4}
\sum_{\widetilde{\pv}_a, \widetilde{\pv}_b,\widetilde{\qv}_a}
  \left(\pv+\frac{\qv}{2}\right) \times \left( \widetilde{\pv}_a +\frac{\widetilde{\qv}_a}{2}\right) \cdot {\boldsymbol \sigma} \check G_{\widetilde{\pv}_a,\widetilde{\qv}_a} \check G_{\widetilde{\pv}_b,\qv-\widetilde{\qv}_a} . \nonumber
\end{eqnarray}
We Fourier transform back with respect to the momentum $\qv$ and  neglect derivatives with respect to $\rv$, i.e. we confine to lowest order in the gradient expansion. We then get
\begin{eqnarray}
\check\Sigma_a^{SS} & = &-n_iv_0^3 \frac{i\lambda_0^2}{4}\sum_{\widetilde{\pv}_a, \widetilde{\pv}_b} \check G(\widetilde{\pv}_a,\rv) \check G(\widetilde{p}_b,\rv) \widetilde{\pv}_b\times p\cdot{\boldsymbol \sigma}  \label{EQ32}\\
\check\Sigma_b^{SS} & = &-n_iv_0^3 \frac{i\lambda_0^2}{4}
\sum_{\widetilde{p}_a, \widetilde{\pv}_b}
 \check G(\widetilde{\pv}_a,\rv) \widetilde{\pv}_a \times \widetilde{\pv}_b\cdot{\boldsymbol \sigma} \check G(\widetilde{\pv}_b,\rv) \label{EQ33}\\
\check\Sigma_c^{SS} &=&-n_iv_0^3\frac{i\lambda_0^2}{4}
\sum_{\widetilde{\pv}_a, \widetilde{\pv}_b}
  \pv\times \widetilde{\pv}_a \cdot {\boldsymbol \sigma} \check G(\widetilde{\pv}_a,\rv) \check G(\widetilde{\pv}_b,\rv) . \label{EQ34}
\end{eqnarray}

When Rashba SOC is present 
one has to consider the covariant self-energy, as done for the side jump and spin current swapping contribution. 
To leading order in the gradient expansion, this is done simply by replacing the Green function $\check G$ 
with its covariant expression $\check{\tilde G}$. Hence the self-energy responsible for the skew scattering reads
\begin{eqnarray}
\check{\tilde \Sigma}_a^{\rm SS} & = & -{\rm i} n_i v_0^3\frac{\lambda_0^2}{4} \sum_{{\bf p}_a,{\bf p}_b} \  \check{\tilde G}({\bf p}_a,\rv) \  \check{\tilde G}({\bf p}_b,\rv)  \ {\bf p}_b \times {\bf p}\cdot {\boldsymbol \sigma} \label{ss5}\\
\check{\tilde \Sigma}_b^{\rm SS} & = &  -{\rm i} n_i v_0^3\frac{\lambda_0^2}{4}\sum_{{\bf p}_a,{\bf p}_b}\  \check{\tilde G}({\bf p}_a,\rv) \ {\bf p}_a \times {\bf p}_b \cdot {\boldsymbol \sigma} \  \check{\tilde G}({\bf p}_b,\rv)  \ \label{ss6}\\ 
\check{\tilde \Sigma}_c^{\rm SS} &=&   -{\rm i} n_i v_0^3\frac{\lambda_0^2}{4} \sum_{{\bf p}_a,{\bf p}_b} \ {\bf p} \times {\bf p}_a \cdot {\boldsymbol \sigma} 
\  \check{\tilde G}({\bf p}_a,\rv) \  \check{\tilde G}({\bf p}_b,\rv) \label{ss7}.
\end{eqnarray}
Since we are considering the effect to first order in $\lambda_0^2$,
the covariant Green functions entering Eqs.(\ref{ss5}-\ref{ss7}) are spin independent and isotropic in momentum space. 
As a result the retarded and advanced components of the above self-energies vanish, 
while the Keldysh component survives only for $\check{\tilde \Sigma}_a^{\rm SS}$ and $\check{\tilde \Sigma}_c^{\rm SS}$. 
Their joint contribution, after recalling that $\sum_{{\bf p}}\tilde G^R({\bf p})=-{\rm i}\pi N_0$, 
leads to an extra term on the right hand side of the Boltzmann equation
\begin{equation}
\label{ss8}
 I^{SS}[f]=-2\pi n_i v_0^2  (v_0 \pi N_0) \frac{\lambda_0^2}{4}\sum_{{\bf p'}}\delta (\epsilon_{\bf p}-\epsilon_{\bf p'}) \left\{   {\bf p'}\times {\bf p}\cdot {\boldsymbol \sigma}, \ f_{\bf p'}\right\}.
\end{equation}

Finally by collecting the collision integrals \eref{eq_89} for spin current swapping, \eref{eq_94} for side jump and \eref{ss8} for skew scattering, the Boltzmann equation \eref{eq_43} reads now
\ber
\tilde \partial_t f_{\pv}&+&{{{{\tilde \nabla}}_{\mathbf{ r}}}}\cdot 
\left[ \frac{\mathbf{ p}}{m} f_{\mathbf{ p}}+
\frac{\lambda_0^2}{8\tau}
\langle \left\{ {\boldsymbol \sigma} \times (\pv'-\pv)  , f_{\pv'} \right\}\rangle  \right]\nn\\
&-&
\frac{e}{2}\left\{ 
\left( {\bf E}+ {{\color{black}{ \frac{\bf p}{m}\times {\bf B}}}}\cdot \nabla_{\mathbf{ p}} \right) ,f_{\mathbf{ p}} \right\}  \nn\\
&=& -\frac{1}{\tau}\left( f_{\pv} -\langle f_{\pv'}\rangle\right)
-(\pi  v_0 N_0)    \frac{\lambda_0^2}{4\tau}\langle \left\{   {\bf p'}\times {\bf p}\cdot {\boldsymbol \sigma}, \ f_{\bf p'}\right\}\rangle\nn\\
&-&i \frac{\lambda_0^2}{4\tau }\langle \left[ {\boldsymbol\sigma}\cdot \pv\times \pv',f_{\pv'}\right]\rangle 
\nn\\
&+&\frac{\lambda_0^2}{8\tau }  \langle  
\left(\Big\{\tilde\nabla_{\rv}( {\boldsymbol\sigma}\times \pv'), f_{\pv'}\Big\}-\Big\{\tilde\nabla_{\rv}( {\boldsymbol\sigma}\times \pv), f_{\pv}\Big\} \right)\rangle
\label{full_boltzmann}
\eer
where, being the scattering  elastic, $f_{\pv}=f(\ep_{\pv}, {\hat \pv})$ and $f_{\pv'}=f(\ep_{\pv}, {\hat \pv'})$ with $\langle \dots \rangle$ indicating the integration over the directions of $\pv'$.
\Eref{full_boltzmann} is the Boltzmann equation valid to first order in the gradient expansion and up to first order in the extrinsic SOC $\lambda_0$. By setting to zero the Rashba SOC and any exchange field, the covariant derivatives only include the standard U(1) electromagnetic field. One can then derive the standard results for the spin Hall effect and spin current swapping
\be
\sigma^{sH}_{sj}=e\frac{\lambda_0^2}{4}n , \ \sigma^{sH}_{ss}=e\frac{\lambda_0^2}{4}n p_F^2  \tau v_0, \ \kappa =\frac{\lambda_0^2p_F^2}{4},
\label{conductivities}
\ee
where $\sigma^{sH}_{sj}$ and $\sigma^{sH}_{ss}$ are the spin Hall conductivities\cite{Raimondi_AnnPhys12} for the side jump and skew scattering contribution and
$\kappa$ is the spin current swapping coefficient\cite{Shenprb14}.
When the Rashba SOC is present, the above equation allows the analysis of the interplay between the intrinsic and extrinsic SOC. However, it turns out that such interplay has some subtle aspects, which have led to the suggestion of an non analytical behavior for vanishing Rashba coupling $\alpha$\cite{TsePRB06,
Hankiwicz_PhaseDiag_PRL08,ChengJPCM}.  In the next section we will consider these aspects in detail and show that, indeed, there is no need to invoke a non-analyticity. Rather, one must take into account the fact that the extrinsic SOC introduces a further spin relaxation mechanism.
\section{The impurity-induced spin-orbit coupling: Elliott-Yafet spin relaxation}
\label{sec6}
\begin{figure}
\begin{center}
\includegraphics[width=2.in]{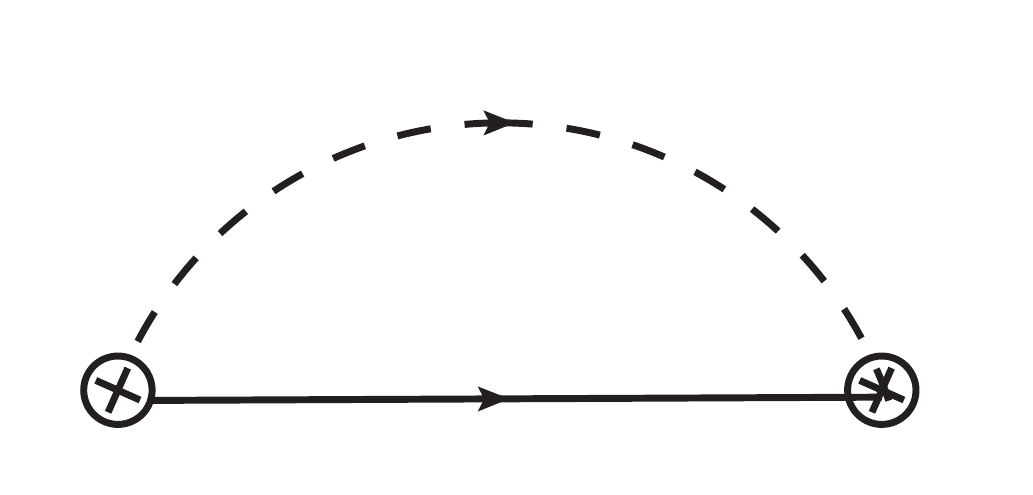}
\caption{Self-energy diagram  in second order in the spin-orbit impurity potential contributing to the Elliott-Yafet spin relaxation.}
\label{fig4}
\end{center}
\end{figure}

The spin-orbit interaction with scattering centers, Eq.~\eqref{eq_70}, gives  rise also  to spin-flip events
leading to the Elliott-Yafet spin relaxation.  Such a process is $\mathcal{O}(\lambda_0^4)$ and shown diagramatically in \fref{fig4}.
Performing the impurity average and defining momenta as in Eq.~\eqref{eq_74}, the Elliott-Yafet self-energy reads 
\ber
\check \Sigma^{EY}(\bp,\bq) &=&
-\frac{\lambda_0^4}{16} n_iv_0^2\sum_{\tilde{\bp}}\,\Big[(\bp+\bq/2)\times(\tilde{\bp}+\bq/2)\Big]\cdot\bsigma
\, \check G_{\tilde{\bp},\bq}\,
\nn\\
&&
\quad\quad\quad
\bsigma\cdot\Big[(\tilde{\bp}-\bq/2)\times(\bp-\bq/2)\Big]
\nn\\
&\approx&
\frac{\lambda^4_0}{16} n_iv_0^2 \sum_{\tilde{\bp}}\,(\bp\times\tilde{\bp})\cdot\bsigma
\, \check G_{\tilde{\bp},\bq}\,\bsigma\cdot(\bp\times\tilde{\bp}).
\eer
In the last line subleading gradient terms $\mathcal{O}(\bq)$ have been neglected.
Considering a 2D system, so that all momenta lie in the $x$-$y$ plane, and transforming back to
the center-of-mass coordinate $\br$, one has (renaming $\tilde{\bp}\rightarrow\bp'$) 
\be
\label{selfEY}
\check{\tilde \Sigma}^{EY}(\bp,\br)=
\frac{\lambda^4_0}{16} n_iv_0^2 \sum_{\bp'}\,\sigma^z
\,\check{\tilde G}(\bp',\br)\,\sigma^z(\bp\times\bp')_z^2.
\ee
Notice that the Elliott-Yafet self-energy is already second order in the (extrinsic) SOC strength,
so that its standard and covariant forms coincide up to 
$\mathcal{O}(\lambda_0^4\,\alpha)$.  
Such higher-order corrections are not needed here, and therefore
covariant quantities were directly introduced in Eq.~\eqref{selfEY}.
The expression \eqref{selfEY} yields the following extra collision term on the right hand side  of the Boltzmann equation \eref{full_boltzmann}:
\be
\label{collisionEY}
{I}^{EY}[f] = -\frac{\lambda_0^4}{16}2\pi n_iv_0^2\sum_{\bp'}\delta(\epsilon_\bp-\epsilon_{\bp'})
\Big[f_\bp-\sigma^z f_{\bp'}\sigma^z\Big](\bp\times\bp')^2_z.
\ee 
The two $\sigma^z$ Pauli matrices flip the in-plane spin components of the distribution function, 
$\sigma^z \Big(f_{\bp'}^{x,y}\sigma^{x,y}\Big) \sigma^z = - f_{\bp'}^{x,y}\sigma^{x,y}$,
while leaving the out-of-plane component $f_{\bp'}^z\sigma^z$ unchanged.  
This is a signature that spin-orbit interaction with the random 2D impurity potential conserves the $z$-spin component,
while relaxing only the in-plane spins -- a result valid strictly in 2D. 
\footnote{One may consider in certain circumstances a more general model including also random Rashba SOC\cite{Dugaev2009a}.}
In order to see this more clearly we follow the procedure of Sec.~\ref{sec3}, and compute
the modification to the spin equation of motion \eqref{rash_3} according to
\be
\frac{1}{2}{\rm Tr} \Big[ \sigma^a \sum_\bp \delta \tilde{I}^{EY}[f] \Big] = -\hat{\Gamma}_{EY}^{ab}s^b,
\ee
where $\hat{\Gamma}_{EY}$ is the Elliott-Yafet spin relaxation matrix.
Explicitly
\ber
\frac{1}{2}{\rm Tr} \Big[ \sigma^{x,y} \sum_\bp \delta \tilde{I}^{EY}[f] \Big]
&=&
-\frac{\lambda_0^4}{16}2\pi n_iv_0^2 \sum_{\bp,\bp'}\delta(\epsilon_\bp-\epsilon_{\bp'})
\Big[f^{x,y}_\bp + f^{x,y}_{\bp'}\Big](\bp\times\bp')^2_z
\nn\\
\label{rateEYxy}
&\approx&
-\frac{1}{\tau}\left(\frac{\lambda_0p_F}{2}\right)^4 s^{x,y}
\eer
and
\ber
\frac{1}{2}{\rm Tr} \Big[ \sigma^z \sum_\bp \delta \tilde{I}^{EY}[f] \Big]
&=&
-\frac{\lambda_0^4}{16}2\pi n_iv_0^2 \sum_{\bp,\bp'}\delta(\epsilon_\bp-\epsilon_{\bp'})
\Big[f^z_\bp - f^z_{\bp'}\Big](\bp\times\bp')^2_z
\nn\\
\label{rateEYz}
&=&
0,
\eer
implying
\be
\label{rateEYfull}
\hat{\Gamma}_{EY} = \frac{1}{\tau_{EY}}
\left(
\begin{array}{ccc}
1 & 0 & 0 \\
0 & 1 & 0 \\
0 & 0 & 0 
\end{array}
\right),\quad
\frac{1}{\tau_{EY}} = \frac{1}{\tau}\left(\frac{\lambda_0p_F}{2}\right)^4.
\ee
In order to obtain Eq.~\eqref{rateEYxy} we employed the diffusive expansion \eqref{eq_46} and set
\be
\langle f_\bp \rangle \approx \frac{1}{N_0} \delta(\epsilon_\bp-\epsilon_F) s^a,
\ee
which is appropriate for the present paramagnetic case.  On the other hand Eq.~\eqref{rateEYz} is obtained
at once by noticing that $(\bp\times\bp')^2_z$ is even under exchange $\bp\leftrightarrow\bp'$.

Elliott-Yafet relaxation, though typically weak, plays a crucial role in the appropriate description
of the spin Hall effect in 2DEGs.  Indeed, the first efforts to combine intrinsic and extrinsic
contributions to the spin Hall effect lead to a puzzling non-analytical behaviour \cite{TsePRB06,
Hankiwicz_PhaseDiag_PRL08,ChengJPCM}:
the spin Hall conductivity for a purely extrinsic sample, with $\alpha=0$, differed from the value
obtained by considering a system with both mechanisms present, where however $\alpha\rightarrow0$.
This unphysical behaviour is cured by Elliott-Yafet processes \cite{Raimondi09}.  To be definite, 
we follow Ref.~\citenum{Raimondi_AnnPhys12} and consider the coupled dynamics
of $J^z_y$ and $s^y$ in the presence of a $x$-pointing electric field (cf. \eref{cony} and \eref{currzy}). The spin current is
\be
\label{jzy}
J^z_y = 2m\alpha D s^y + (\sigma^{sH}_{int} + \sigma^{sH}_{ext})E_x,
\ee
with $\sigma^{sH}_{ext}=\sigma^{sH}_{sj}+\sigma^{sH}_{ss}$,
while from Eqs.~\eqref{rash_3} and \eqref{rateEYfull} one has for $s^y$
\be
\label{sy}
\partial_t s^y = -\frac{1}{\tau_{EY}} - 2m\alpha  J^z_y.
\ee
Solving Eqs.~\eqref{jzy}-\eqref{sy} yields\cite{Raimondi_AnnPhys12}
\ber
\label{jzyfull}
J^z_y &=& \frac{1}{1+\tau_{EY}/\tau_{DP}}(\sigma^{sH}_{int} + \sigma^{sH}_{ext})E_x
\\
\label{syfull}
s^y &=& -\frac{2m\alpha\tau_{EY}}{1+\tau_{EY}/\tau_{DP}}(\sigma^{sH}_{int} + \sigma^{sH}_{ext})E_x,
\eer
which are respectively the generalisation of the spin Hall 
and inverse spin galvanic/Edelstein effects in the presence of intrinsic (Rashba) and extrinsic spin-orbit coupling.  
These expressions are analytical and reduce to the known results for either $\alpha\rightarrow0$ or $\lambda_0^2\rightarrow0$.
Physically, they show that the behaviour of both effects is determined by
the ratio between Dyakonov-Perel and Elliott-Yafet spin relaxation.

\section{Conclusions}
\label{sec_conclusions}

We have employed the Keldysh formalism -- in its semiclassical limit --
to describe the spin-charge coupled dynamics in a 2DEG with both intrinsic (Rashba) and extrinsic
sources of spin-orbit interaction.  Such dynamics are rich and typically rather intricated,
but we have seen that rewriting spin-orbit coupling in terms of non-Abelian gauge fields leads to a compact 
and physically transparent set of equations.  In particular the latter, obtained from the equations of motion of the
locally $SU(2)$-covariant Keldysh Green's function $\check{\TG}(1,2)$, show that:
\begin{itemize}
 \item Spin and charge are coupled via the $SU(2)$ field tensor (``spin-electric'' and ``spin-magnetic'' fields);
 \item The spin obeys an $SU(2)$-covariant continuity equation, appropriately modified when extrinsic spin-orbit is present;
 this corrects both the definition of the spin current and the collision integral, but preserves their covariance properties;
 \item The side-jump mechanism is naturally seen as a modification of the velocity operator arising from the 
 $\bq\neq0$ corrections to the Born self-energy;
 \item Skew scattering and spin current swapping are interference processes proportional to ${\rm Re} (AB^*)$ and ${\rm Im} (AB^*)$, respectively, where $A$ and $B$ are the scattering amplitudes defined in \eref{eq_71}.   Skew scattering 
 arises  beyond the Born approximation, which is instead enough to have the spin current swapping.
 Both processes, in contrast with side jump,
are  due to $\bq=0$ self-energy terms;
 \item Elliott-Yafet spin relaxation introduces a typically small though crucial energy scale, 
 which is necessary to cure unphysical non-analytical behaviours of various physical quantities. 
\end{itemize}
The Keldysh non-Abelian approach we have outlined focusing on the paradigmatic Rashba 2DEG
can actually be (and indeed has been) employed in a wide range of systems, and offers certain further advantages.
Let us briefly mention a few.
\begin{itemize}
 \item The lack of spin conservation in the presence of spin-orbit coupling may lead to ambiguous definitions of, e.g. spin currents.
 This was extensively debated from early on \cite{Rashba2003} and posed problems concerning Onsager
 reciprocity.\cite{wang2012}  Such problems were solved via the non-Abelian formulation,\cite{gorini2012,ShenVR14}
 by construction devoid of any ambiguity.\cite{tokatly2008}
 \item The kinetic equations can include any linear-in-momentum spin-orbit field, e.g. \`{a} la Rashba-Dresselhaus.  Furthermore,
 (pseudo)spin-orbit coupling in $N$-band models can be written in terms of $SU(N)$ gauge fields, and formally handled 
exactly as we did in the single band, 2$\times$2 case.
 \item The formalism can describe the dynamics of exotic systems, such as cold atoms in artificial gauge fields.\cite{tokatly2016}
 \item Non-homogeneous and/or dynamical spin-orbit coupling, e.g. gate-controlled \cite{brouwer2002, malshukov2003, GoriniPRB10} 
 or due to thermal vibrations \cite{gorini2015}, has numerous practical and theoretical implications, and is by construction
 included in the non-Abelian approach.  In a similar way, the latter can deal with the spin and charge dynamics induced by time-dependent
 magnetic textures.\cite{tserkovnyak2008} 
\end{itemize}


\begin{thebibliography}{10}

\bibitem{GoriniPRB10}
C.~Gorini, P.~Schwab, R.~Raimondi and A.~L. Shelankov, {Non-Abelian gauge
  fields in the gradient expansion: Generalized Boltzmann and Eilenberger
  equations}, {\em Phys. Rev. B} {\bf 82}, p. 195316 (Nov 2010).

\bibitem{Raimondi_AnnPhys12}
R.~Raimondi, P.~Schwab, C.~Gorini and G.~Vignale, {Spin-orbit interaction in a
  two-dimensional electron gas: a SU(2) formulation}, {\em Ann. Phys.} {\bf
  524}, p. 153  (2012).
  
 \bibitem{Tatara2016} G. Tatara, Theory of electron transport and magnetization dynamics in
metallic ferromagnets, in this lecture series (2017).

\bibitem{Rammer_qkt}
J.~Rammer, {\em {Quantum Field Theory of Nonequilibrium States}} (Cambridge
  University Press, Cambridge, 2007).

\bibitem{AGD1975}
A.~A. Abrikosov, L.~P. Gorkov and I.~E. Dzyaloshinski, {\em {Methods of Quantum
  Field Theory in Statistical Physics}} (Dover Publications, 1975).

\bibitem{Rashba84}
Y.~A. Bychkov and E.~I. Rashba, {Oscillatory effects and the magnetic
  susceptibility of carriers in inversion layers}, {\em J. Phys. C} {\bf 17},
  p. 6039  (1984).

\bibitem{Edelstein90}
V.~M. Edelstein, {}, {\em Solid State Commun.} {\bf 73}, p. 233  (1990).

\bibitem{Aronov89}
A.~G. Aronov and Y.~B. Lyanda-Geller, {\em JETP Lett.} {\bf 50}, p. 431
  (1989).

\bibitem{Ganichev02}
S.~D. Ganichev, E.~L. Ivchenko, V.~V. Belkov, S.~A. Tarasenko, M.~Sollinger,
  D.~Weiss, W.~Wegscheider and W.~Prettl, {Spin-galvanic effect}, {\em Nature}
  {\bf 417}, p. 153  (2002).

\bibitem{Ganichev2016}
S.~D. {Ganichev}, M.~{Trushin} and J.~{Schliemann}, {Spin polarisation by
  current}, {\em ArXiv e-prints}  (June 2016).

\bibitem{Raimondi05}
R.~Raimondi and P.~Schwab, {}, {\em Phys. Rev. B} {\bf 71}, p. 033311  (2005).

\bibitem{toelle2014}
S.~T{\"o}lle, C.~Gorini and U.~Eckern, {Room-temperature spin thermoelectrics
  in metallic films}, {\em Phys. Rev. B} {\bf 90}, p. 235117 (Dec 2014).

\bibitem{Engel05}
H.-A. Engel, B.~I. Halperin and E.~Rashba, {Theory of Spin Hall Conductivity in
  n-Doped {GaAs}}, {\em Phys. Rev. Lett.} {\bf 95}, p. 166605  (2005).

\bibitem{Tse06}
W.-K. Tse and S.~{Das Sarma}, Spin hall effect in doped semiconductor
  structures  (2006).

\bibitem{Lifshitz2009}
M.~B. Lifshits and M.~I. Dyakonov, Swapping spin currents: Interchanging spin
  and flow directions, {\em Phys. Rev. Lett.} {\bf 103}, p. 186601 (Oct 2009).

\bibitem{Berger1970}
L.~Berger, Side-jump mechanism for the hall effect of ferromagnets, {\em Phys.
  Rev. B} {\bf 2}, 4559 (Dec 1970).

\bibitem{Shenprb14}
K.~Shen, R.~Raimondi and G.~Vignale, {Theory of coupled spin-charge transport
  due to spin-orbit interaction in inhomogeneous two-dimensional electron
  liquids}, {\em Phys. Rev. B} {\bf 90}, p. 245302 (Dec 2014).

\bibitem{TsePRB06}
W.-K. Tse and S.~{Das Sarma}, {}, {\em Phys. Rev. B} {\bf 74}, p. 245309
  (2006).

\bibitem{Hankiwicz_PhaseDiag_PRL08}
E.~M. Hankiewicz and G.~Vignale, {Phase Diagram of the Spin Hall Effect}, {\em
  Phys. Rev. Lett.} {\bf 100}, p. 026602  (2008).

\bibitem{ChengJPCM}
J.~L. Cheng and M.~W. Wu, {Kinetic investigation of the extrinsic spin Hall
  effect induced by skew scattering}, {\em Journal of Physics: Condensed
  Matter} {\bf 20}, p. 085209  (2008).

\bibitem{Dugaev2009a}
V.~K. Dugaev, E.~Y. Sherman, V.~I. Ivanov and J.~Barna\ifmmode~\acute{s}\else
  \'{s}\fi{}, Spin relaxation and combined resonance in two-dimensional
  electron systems with spin-orbit disorder, {\em Phys. Rev. B} {\bf 80}, p.
  081301 (Aug 2009).

\bibitem{Raimondi09}
R.~Raimondi and P.~Schwab, {Tuning the Spin Hall Effect in a Two-Dimensional
  Electron Gas}, {\em Europhys. Lett.} {\bf 87}, p. 37008  (2009).

\bibitem{Rashba2003}
E.~I. Rashba, {Spin currents in thermodynamic equilibrium: The challenge of
  discerning transport currents}, {\em Phys. Rev. B} {\bf 68}, p. 241315(R)
  (2003).

\bibitem{wang2012}
L.~Y. Wang, A.~G. Mal'shukov and C.~S. Chu, {Nonuniversality of the intrinsic
  inverse spin-Hall effect in diffusive systems}, {\em Phys. Rev. B} {\bf 85},
  p. 165201  (2012).

\bibitem{gorini2012}
C.~Gorini, R.~Raimondi and P.~Schwab, {Onsager Relations in a Two-Dimensional
  Electron Gas with Spin-Orbit Coupling}, {\em Phys. Rev. Lett.} {\bf 109}, p.
  246604  (2012).

\bibitem{ShenVR14}
K.~Shen, G.~Vignale and R.~Raimondi, {Microscopic Theory of the Inverse
  Edelstein Effect}, {\em Phys. Rev. Lett.} {\bf 112}, p. 096601 (Mar 2014).

\bibitem{tokatly2008}
I.~V. Tokatly, {Equilibrium Spin Currents: Non-Abelian Gauge Invariance and
  Color Diamagnetism in Condensed Matter}, {\em Phys. Rev. Lett.} {\bf 101}, p.
  106601  (2008).

\bibitem{tokatly2016}
I.~V. Tokatly and E.~Y. Sherman, {Spin evolution of cold atomic gases in
  $SU(2)\otimes U(1)$ fields}, {\em Phys. Rev. A} {\bf 93}, p. 063635  (2016).

\bibitem{brouwer2002}
P.~W. Brouwer, J.~N. H.~J. Cremers and B.~I. Halperin, {Weak localization and
  conductance fluctuations of a chaotic quantum dot with tunable spin-orbit
  coupling}, {\em Phys. Rev. B} {\bf 65}, p. 081302(R)  (2002).

\bibitem{malshukov2003}
A.~G. Mal'shukov, C.~S. Tang, C.~S. Chu and K.~A. Chao, {Spin-current
  generation and detection in the presence of an ac gate}, {\em Phys. Rev. B}
  {\bf 68}, p. 233307  (2003).

\bibitem{gorini2015}
C.~Gorini, U.~Eckern and R.~Raimondi, {Spin Hall Effects Due to Phonon Skew
  Scattering}, {\em Phys. Rev. Lett.} {\bf 115}, p. 076602  (2015).

\bibitem{tserkovnyak2008}
Y.~Tserkovnyak and M.~Mecklenburg, {Electron transport driven by nonequilibrium
  magnetic textures}, {\em Phys. Rev. B} {\bf 77}, p. 134407  (2008).

\end{thebibliography}

\end{document}